\documentclass{iopart}
\usepackage{graphicx}
\usepackage{amssymb}

\begin{document}

%%% TITLE, AUTHORS, ABSTRACT
\title[Driven Vortex Dynamics in Systems with Two-Fold Anisotropy...]{Driven Superconducting Vortex Dynamics in Systems with Two-Fold Anisotropy in the Presence of Pinning}

\author{E.~Roe$^1$, M.~R.~Eskildsen$^1$, C.~Reichhardt$^2$, and C.~J.~O.~Reichhardt$^2$}
\address{$^1$ Department of Physics, University of Notre Dame, Notre Dame, Indiana 46656, USA}
\address{$^2$ Theoretical Division, Los Alamos National Laboratory, Los Alamos, New Mexico 87545, USA}
\ead{cjrx@lanl.gov}

%%% ABSTRACT
\begin{abstract}
We examine the dynamics of superconducting vortices with two-fold anisotropic interaction potentials driven over random pinning and compare the behavior under drives applied parallel and perpendicular to the anisotropy direction. The number of topological defects reaches a maximum near depinning and then drops with increasing driving force as the vortices form one-dimensional chains. This coincides with a transition from a pinned nematic to a moving smectic aligned with the soft direction of the anisotropy. The system is generally more ordered when the drive is applied along the soft direction of the anisotropy, while for driving along the hard direction, there is a critical value of the anisotropy above which the system remains aligned with the soft direction. We also observe hysteresis in the dynamics, with one-dimensional aligned chains persisting during a decreasing drive sweep to drives below the threshold for chain formation during the increasing drive sweep. More anisotropic systems have a greater amount of structural disorder in the moving state. For lower anisotropy, the system forms a moving smectic-A state, while at higher anisotropy, a moving nematic state appears instead. 
\end{abstract}

\maketitle

\vskip 2pc

%%% INTRO
\section{Introduction}

A wide range of systems can be described
effectively as an assembly of particles interacting with each other
and with quenched disorder,
leading to the appearance of
depinning and multiple sliding phases \cite{Fisher98,Reichhardt17}.
Such systems include
vortices in type-II superconductors \cite{Blatter94,Reichhardt17},
colloidal particles \cite{Reichhardt02,Pertsinidis08,Tierno12},
active matter \cite{Sandor17,Morin17},
magnetic skyrmions \cite{Jiang17,Reichhardt15a}, 
pattern forming systems \cite{Reichhardt03,Zhao13}, and 
sliding Wigner crystals \cite{Williams91,Reichhardt01}.
In each case there is a threshold for
motion or a depinning transition
above which the particles can depin into a disordered or fluctuating state
containing
numerous topological defects \cite{Bhattacharya93,Reichhardt17}.
One of the most
studied systems of this type is vortices in type-II superconductors,
which can exhibit dynamical transitions into more ordered moving phases such 
as a moving crystal \cite{Koshelev94}, anisotropic crystal \cite{Giamarchi96},
or moving smectic \cite{Moon96,Balents98,Pardo98,Olson98a}.
These
transitions are associated with changes in the 
structure factor \cite{Giamarchi96,Moon96,Balents98,Pardo98,Olson98a}, 
the number and
orientation of topological defects \cite{Moon96,Pardo98,Olson98a}, 
and the noise characteristics
\cite{Marley95,Olson98a,Kolton99,Diaz17,Sato19}. 
In a two-dimensional (2D)
system driven over random disorder,
the fluctuations experienced by the particle due to its motion
over the quenched disorder are
anisotropic, leading to the formation of a
moving smectic state  \cite{Balents98}.
Beyond superconducting vortices, moving smectics
have also been studied 
in other 2D systems driven over quenched disorder
including Wigner crystals \cite{Reichhardt01}
and frictional systems \cite{Granato11}. 

In most of these systems,
the particle-particle interactions are isotropic,
so in the absence of quenched
disorder an isotropic crystal appears. For example,
in the case of type-II superconductors with isotropic 
repulsion, the vortices form a triangular lattice \cite{Blatter94}.
There are, however, many examples of particle-like systems
that have two-fold anisotropic interactions, 
including colloidal particles in tilted magnetic fields
\cite{Eisenmann04,Froltsov05},
dusty plasmas \cite{Yang19}, 
electron liquid crystal states
\cite{Kivelson98,Lilly99a,Fu20},
skyrmions \cite{Lin15,Wang17a,Nagase19}, 
and superconducting vortices
\cite{Blatter91b,Balents95,Gordeev00,Carlson03,Nie05,Reichhardt06a,Shibata15,delValle15}. 
Anisotropic vortex-vortex interactions can arise from
anisotropy in the material or nematicity in the substrate, or it can be
induced by a tilted field.
Theoretical work on vortex liquid crystal systems with two-fold
anisotropy showed that these systems can 
form smectic-A states and exhibit two step melting transitions
\cite{Carlson03,Reichhardt06a}.
Magnetic skyrmions
have many similarities to
superconducting vortices and typically form 
a triangular lattice under
isotropic conditions \cite{Muhlbauer09,Yu10}.
Two-dimensional anisotropic skyrmions
can produce what are called skyrmion
liquid crystals
with smectic \cite{Nagase19} or more specifically smectic-A
ordering \cite{Azaroff80}.
Far less is known about the behavior of driven states
of anisotropic crystals in quenched disorder,
such as what dynamical ordering transitions appear
and what differences arise when the driving is applied
parallel or perpendicular to the anisotropy direction.
In this work we study the dynamics of superconducting vortices with a
twofold anisotropy potential that are driven over
quenched disorder
both parallel and perpendicular to the anisotropy
for varied values of the anisotropy and the quenched disorder strength.
In addition to superconducting vortices,
our results should also be relevant to the wider class of assemblies
of particles with twofold anisotropic interactions
moving over quenched disorder.   

%%% METHODS
\section{Methods}
\label{Methods}
In previous computational modeling of vortices as point particles
interacting with pinning the
vortices had a pairwise isotropic repulsive potential that is proportional
to the zeroth order Bessel function, $U(r) = K_0(r)$ \cite{Tinkham:1996un},
causing the vortices to form a triangular ground state in
the absence of quenched disorder. 
Point particle models of vortices with
two-fold, four-fold, and six-fold anisotropic interactions 
have also been considered
\cite{Olszewski:2018fp,Olszewski20}
in the context of triangular to square and other rotational transitions. 
In anisotropic systems with
a total of $n_a$ anisotropy axes,
the vortex-vortex interaction potential
has the form
\begin{equation}
  U(r,\theta) = A_v K_0(r)\left[ 1 + K_a \cos^2 \left( \frac{n_a (\theta-\phi_a)}{2} \right)\right] \label{potentialeqn}
\end{equation}
where $r = |{\bf r}_i - {\bf r}_j|$ is the distance between two 
vortices at positions ${\bf r}_i$ and ${\bf r}_j$ and the angle
between the vortices with respect to the $x$-axis is
 $\theta = \tan^{-1}(r_y / r_x)$ with ${\bf r} = {\bf r}_i - {\bf r}_j$, $r_x = {\bf r} \cdot {\bf \hat{x}}$ and $r_y = {\bf r} \cdot {\bf \hat{y}}$.
Here $A_v$ is an isotropic vortex interaction
strength that we use as a normalization parameter.
The magnitude of the anisotropic contribution to the vortex interaction 
is given by $K_a$. 
In this work we consider two-fold anisotropy ($n_a = 2$)  or $K_{2}$.
Specifically, we
examine the dynamics of systems with different
anisotropy strengths driven over quenched disorder.

\begin{figure*}
  \begin{center}
\includegraphics[width=\textwidth]{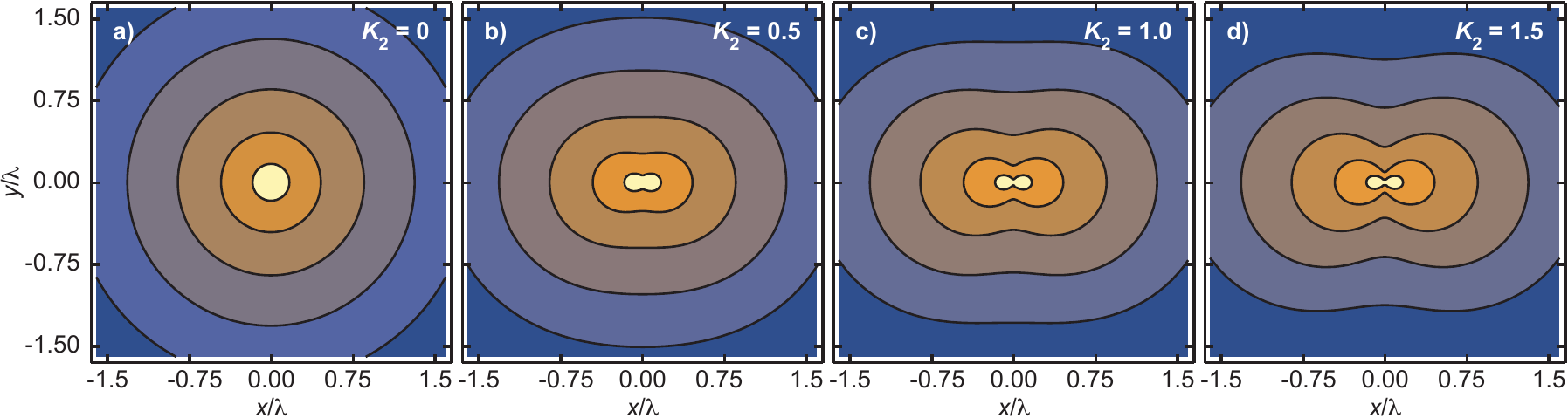}
  \end{center}
    \caption{
      Equipotential lines for the vortex-vortex interaction in Eq. (\ref{potentialeqn}) for increasing anisotropy parameter $K_{2}=0.0$ (a), 0.5 (b), 1.0 (c), and
      1.5 (d).
      The deep pinch points along the $y$-axis cause chain states to form that affect the dynamics of the system.}
    \label{fig:1}
\end{figure*}

In Fig.~\ref{fig:1} we illustrate the equipotential
lines for the two-fold anisotropic vortex potential
with $K_{a} = 0.0$, 0.5, 1.0, and 1.5.
For $K_{2} = 0.0$ in Fig.~\ref{fig:1}(a),
the interaction is isotropic and the vortices form a triangular lattice.
As the anisotropy increases, the potentials become more elongated, 
implying that the vortex-vortex interaction forces are strongest along
the $x$-direction and weaker along the $y$-direction.
In previous work, anisotropy was introduced by multiplying the force
components of an isotropic interaction potential by different factors in
the $x$- and $y$-directions
\cite{Reichhardt06a}. 
Although this approach produced anisotropic diffusion,
smectic ordering appeared only for very large differences 
in the multiplication prefactors,
making this an unrealistic representation of anisotropic systems.  
A much better realization is the
2D anisotropic potential of the type shown in Fig.~\ref{fig:1},
which produces a much more complicated force configuration than a simple
multiplication factor would.

We consider a two-dimensional system of size $L \times L$
with $L = 72\lambda$, where $\lambda$ is the London penetration depth. 
The dynamics of vortex $i$ are governed by the
following overdamped equation of motion:
\begin{equation}
  \eta \frac{d{\bf r}_i}{dt} = {\bf F}_i^{vv} + {\bf F}_i^T+{\bf F}_i^p.
\end{equation}
Here $\eta$ is the damping constant, which is set to unity.
The vortex-vortex
interaction force is ${\bf F}^{vv} = -{\bf \nabla}(U) = (-\partial U/\partial x, -\partial U/\partial y)$.
With a twofold anisotropy and $ \phi_a=0$, the force is 
%\begin{widetext}
%  \begin{subequations}
    \begin{eqnarray}
      F_x&=A_v \left [ \cos(\theta) K_1(r) \left ( 1 + K_2 \cos^2 (\theta) \right ) - \frac{K_2}{r} K_0(r) \sin(\theta)\sin (2\theta) \right
      ] \\
      F_y&=A_v\left[ \sin(\theta)K_1(r)\left ( 1+ K_2 \cos^2 (\theta) \right ) + \frac{K_2}{r}K_0(r) \cos(\theta)\sin (2\theta) \right]
    \end{eqnarray}
%  \end{subequations}
    %\end{widetext}
    There are a total of $N_v$ vortices in the sample.
Each vortex also experiences forces 
${\bf F}_i^p$ from the substrate, which is modeled
as $N_p$ parabolic pinning traps placed in random but non-overlapping positions. 
Each pinning site is of radius $r_p$ 
and can exert a maximum force of $F_p$.
The vortex-pin interaction is directed toward the center of the pinning site and is given by
$  {\bf F}_i^p = F_p \sum_j^{N_p} ({\bf r}_j^p - {\bf r}_i) \, \Theta(r_p - |{\bf r}_i - {\bf r}_j^p|){\hat {\bf r}_{ij}^{(p)}}$. 
For the parameters we
consider, an individual pinning site can capture at most one vortex.   
In this work we fix $r_p = 0.5\lambda$
and $F_{p} = 0.5$. 

The initial vortex positions are
obtained using simulated annealing.
Starting from a high temperature where the vortices are in a liquid state,
we lower the temperature to zero in a series of steps.
Thermal forces arise from Langevin kicks ${\bf F}_i^T$ 
with the properties $\langle {\bf F}^T \rangle = 0.0$ and
$\langle {\bf F}_i^T(t) \, {\bf F}_j^T(t') \rangle = 2 \eta k_B T \delta_{ij} \, \delta(t-t')$ where $k_B$ is the Boltzmann constant.
We begin the annealing process at $F^T = 4.0$
where the vortices are rapidly diffusing,
and gradually cool the system to $F^T =0.0$.
The temperature is reduced by $\Delta F^T = -0.05$
every $10^{4}$ simulation time steps. 

After annealing we apply a driving force in either the $x$- or $y$-direction.
Here $x$ is the hard direction of the anisotropy
along which the vortices are more repulsive,
and $y$ is the soft direction. 
We start at $F_{D}=0.0$ and increase the force in increments of
$\Delta F_{D} = 0.05$ every $10^{4}$
simulation time steps up to a maximum drive of $F_{\rm max}=1.5$.
We then decrease the drive by $\Delta F_{D}$ every
$10^{4}$ time steps until $F_{D} =0.0$ again. 
We use a pinning density $\rho_p=N_p/L^2$ ranging
from $\rho=0$ to $\rho=0.48225$ and fix the vortex
density $\rho_v=N_v/L^2$ to $\rho_v=0.44$.
For all results discussed here, values are obtained by averaging the results of 10 simulations for each set of parameters.

Increasing or decreasing the value of $K_2$ changes the magnitude of the energy potential experienced by each vortex, which is equivalent to a change in the effective vortex density.
To eliminate effects arising from a density difference, we define an effective magnetic field that is proportional to the two-dimensional integral of the interaction potential:
\begin{equation}
  B_{\rm{eff}} \propto \int_0^{\infty} r \, dr \int_0^{2\pi} d\theta \; U(r,\theta) \propto A_v \left( 2+K_2 \right).
  \label{Beff}
\end{equation}
Using $K_2 = 0$ and $A_v = 2.0$ as a reference, we set $A_v = 4/(2 + K_2)$ for each individual molecular dynamics simulation, such that all simulations have the same value of $B_{\rm{eff}}$.
The vortex lattice configurations are analyzed after  annealing and during the drive sweep processes by calculating the structure factor, $S({\bf k})$, and by 
using a Voronoi polygon construction.
This yields the local coordination number $z_i$ of each vortex, which is used to compute the fractions
$P_n = \frac{1}{N_v} \sum_{i=1}^{N_v} \delta(z_i - n)$ for $n = 5$, 6, and 7. The most useful parameter is the fraction of defects $P_d=1-P_6$,
which provides a measure of the disorder of the system. 

\section{Results}
\begin{figure*}
  \centering
    \includegraphics[width=\textwidth]{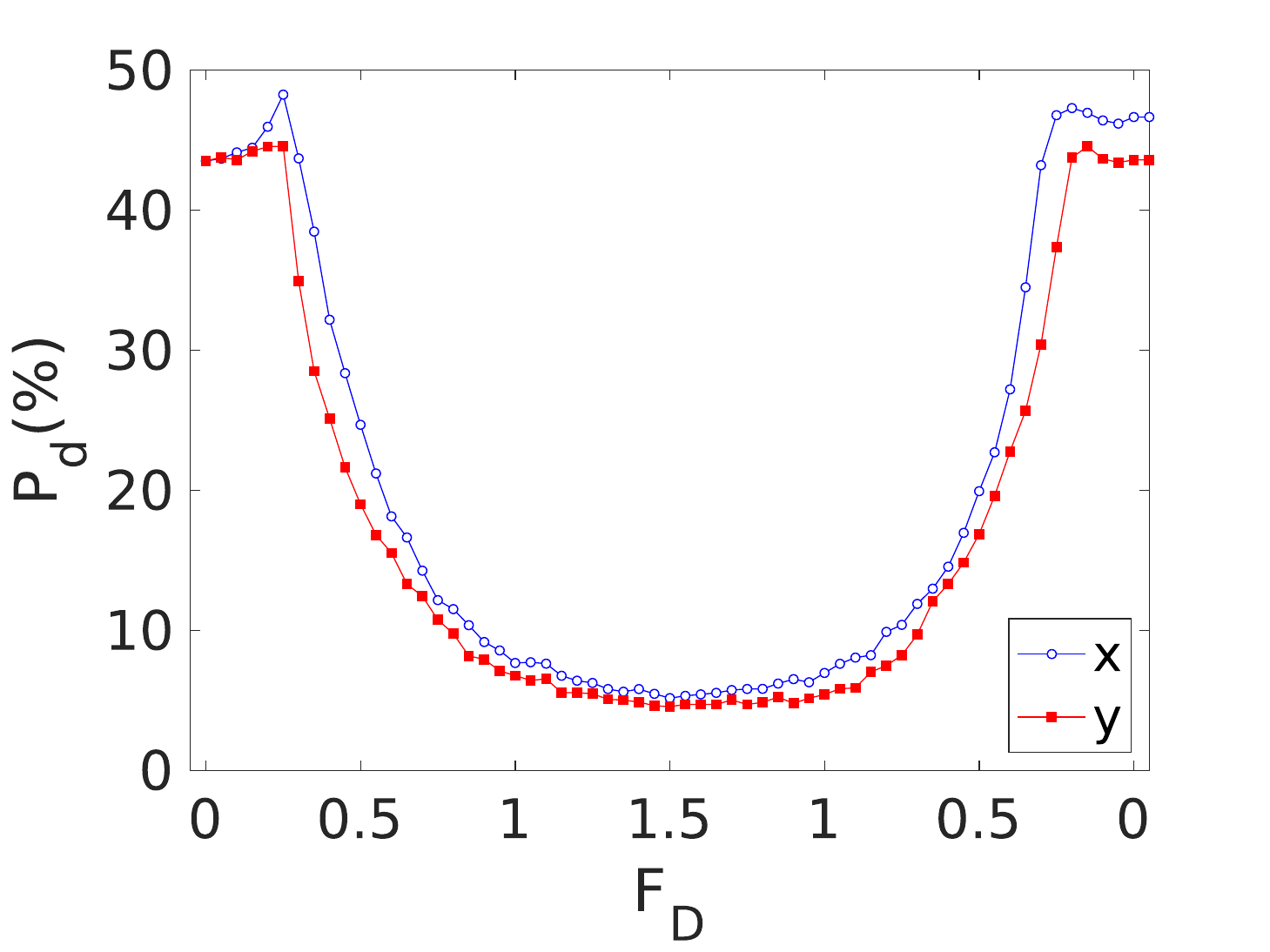}
    \caption{
The fraction of topological defects $P_{d}$ versus the driving force $F_{D}$
in a system with $K_2 = 0.55$ and $\rho_p = 0.48225$ for $x$-direction
driving (blue) and $y$-direction driving (red).
    }
\label{fig:2}
\end{figure*}

In Fig.~\ref{fig:2} we plot the fraction of topological defects $P_{d}$
versus driving force $F_{D}$ 
for a system with $K_2 = 0.55$ and $\rho_p = 0.48225$
under driving in the $x$- and $y$-directions.
The system is in a disordered configuration after the annealing
process with a large fraction, $P_d=0.43$, of vortices that are not
sixfold coordinated.
As
$F_{D}$ increases, a depinning transition occurs
near $F_{D} = 0.2$
that coincides with a maximum in $P_d$ of
$P_{d} = 0.45$ for $y$-direction driving and
$P_d=0.47$ for $x$-direction driving.
The depinning threshold
is slightly higher for driving in the $x$-direction.
Above depinning, $P_{d}$ 
rapidly drops
with increasing $F_D$ and approaches
$P_d=0.05$ for $F_{D} >1.0$.
The minimum value of $P_{d}$  
is similar for driving in either direction at this value of $K_{2}$.
For driving in the $y$-direction, the decrease in $P_d$ as $F_D$ increases
is correlated with dynamical ordering into a moving smectic phase
containing a small number
of dislocations
that are aligned in the driving direction,
as observed previously \cite{Moon96,Pardo98,Olson98a}.
One distinction we find
in the anisotropic system is
that the smectic state for driving in the $x$-direction
is {\it not} aligned with the driving direction but is instead aligned
with the soft anisotropy or $y$-direction.

\begin{figure*}
  \centering
  \begin{minipage}{0.7\textwidth}
    \includegraphics[width=0.5\textwidth]{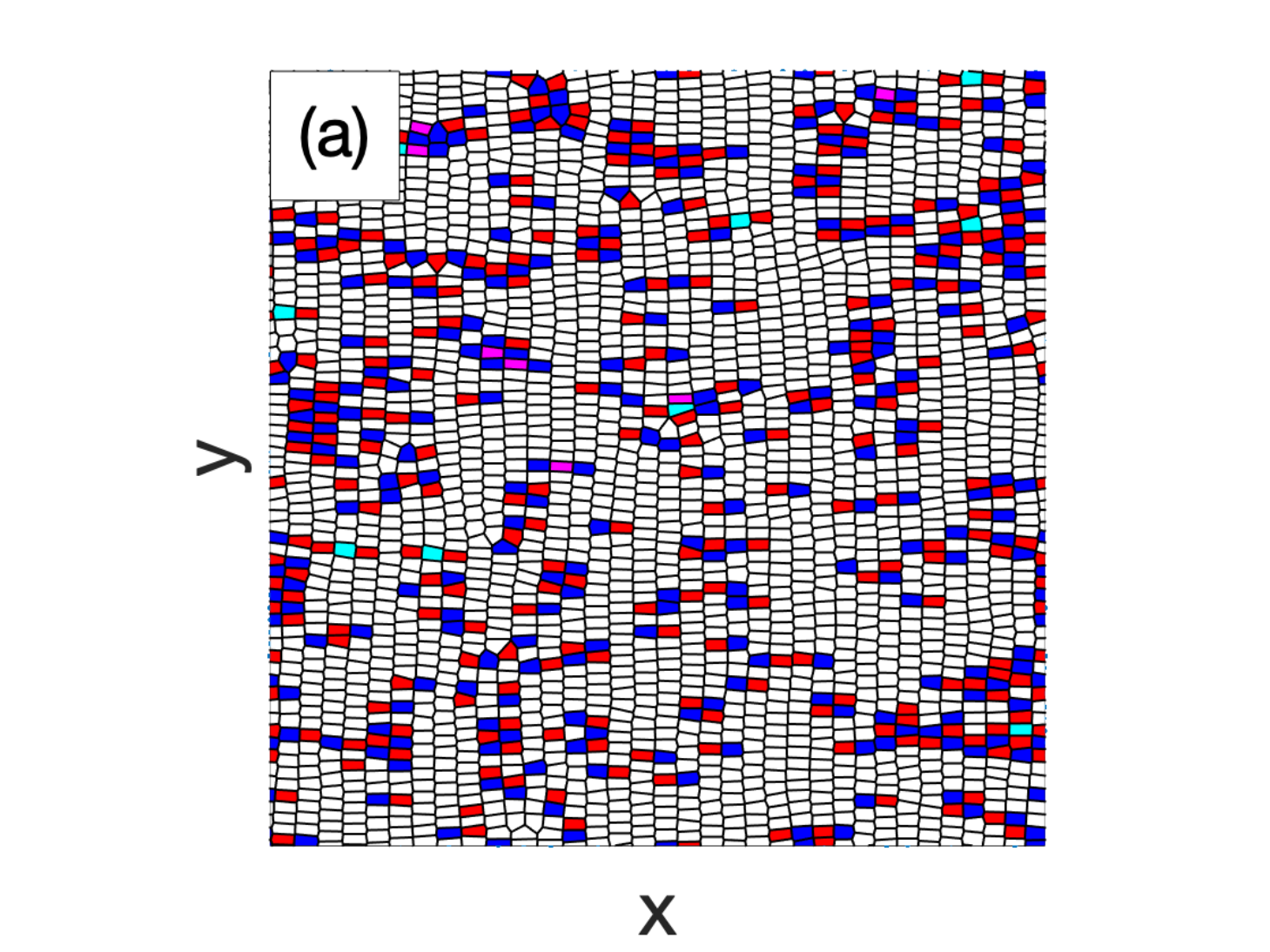}%
    \includegraphics[width=0.5\textwidth]{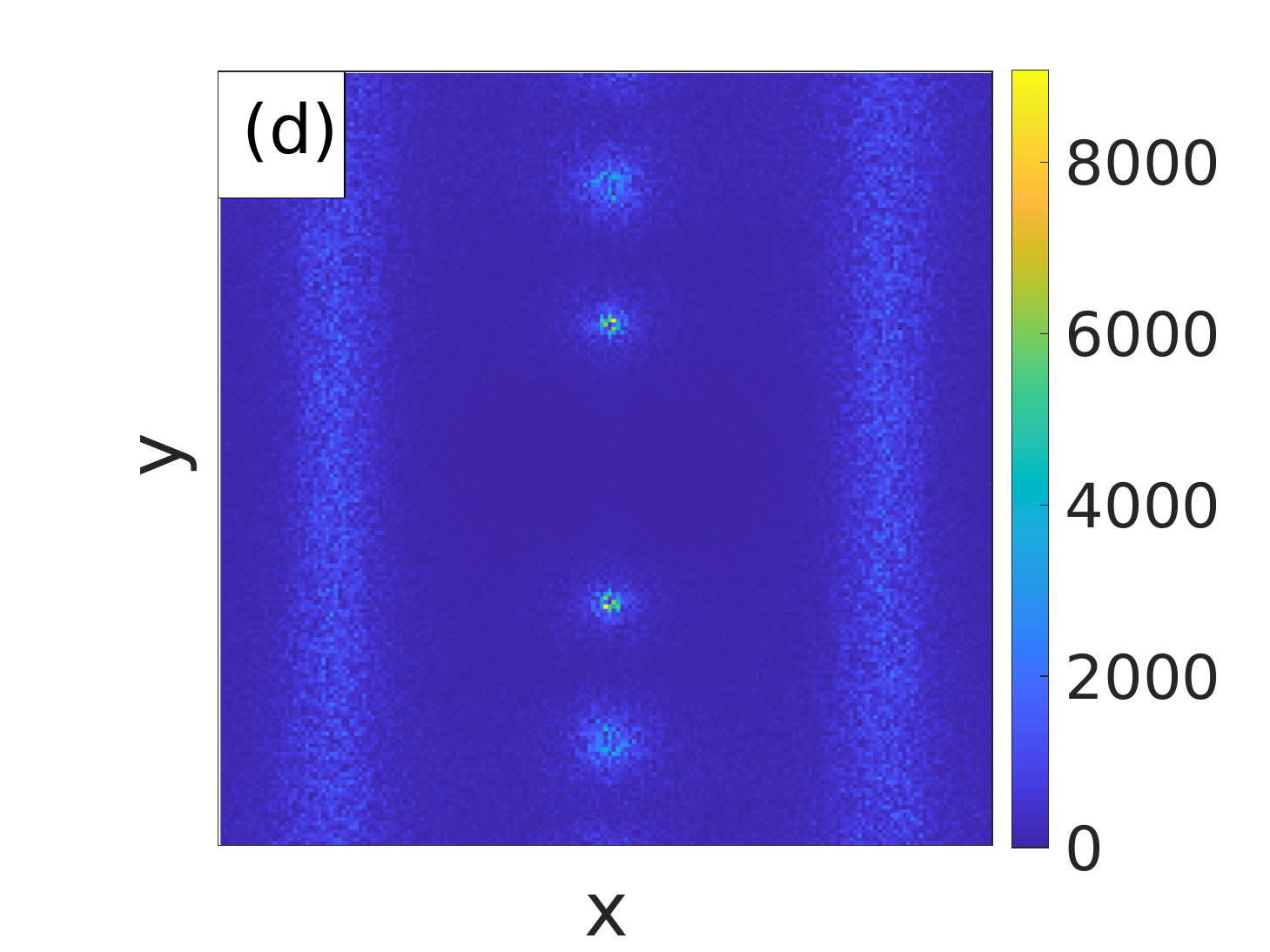}
  \end{minipage}\hfill\\
  \begin{minipage}{0.7\textwidth}
    \includegraphics[width=0.5\textwidth]{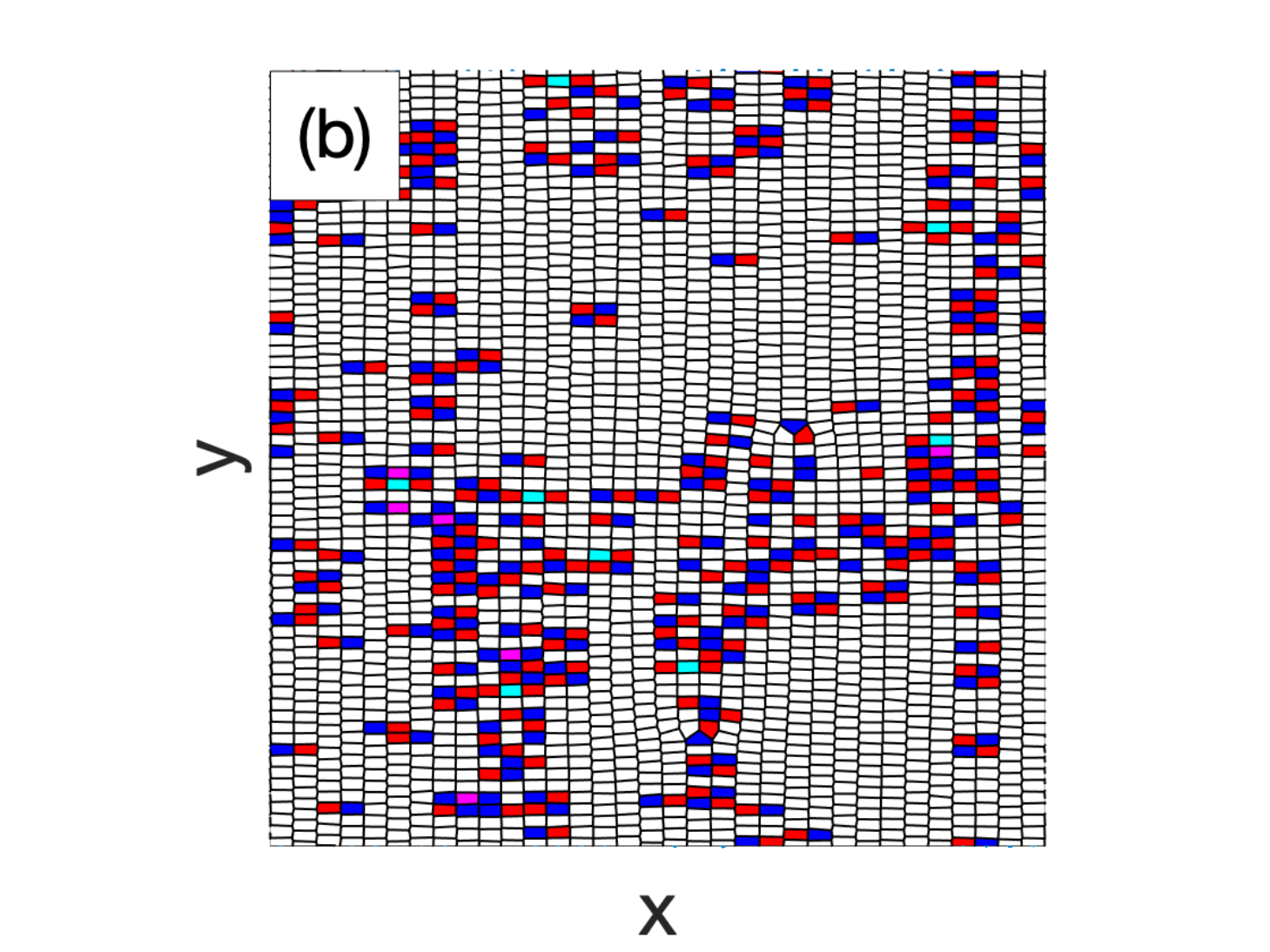}%
    \includegraphics[width=0.5\textwidth]{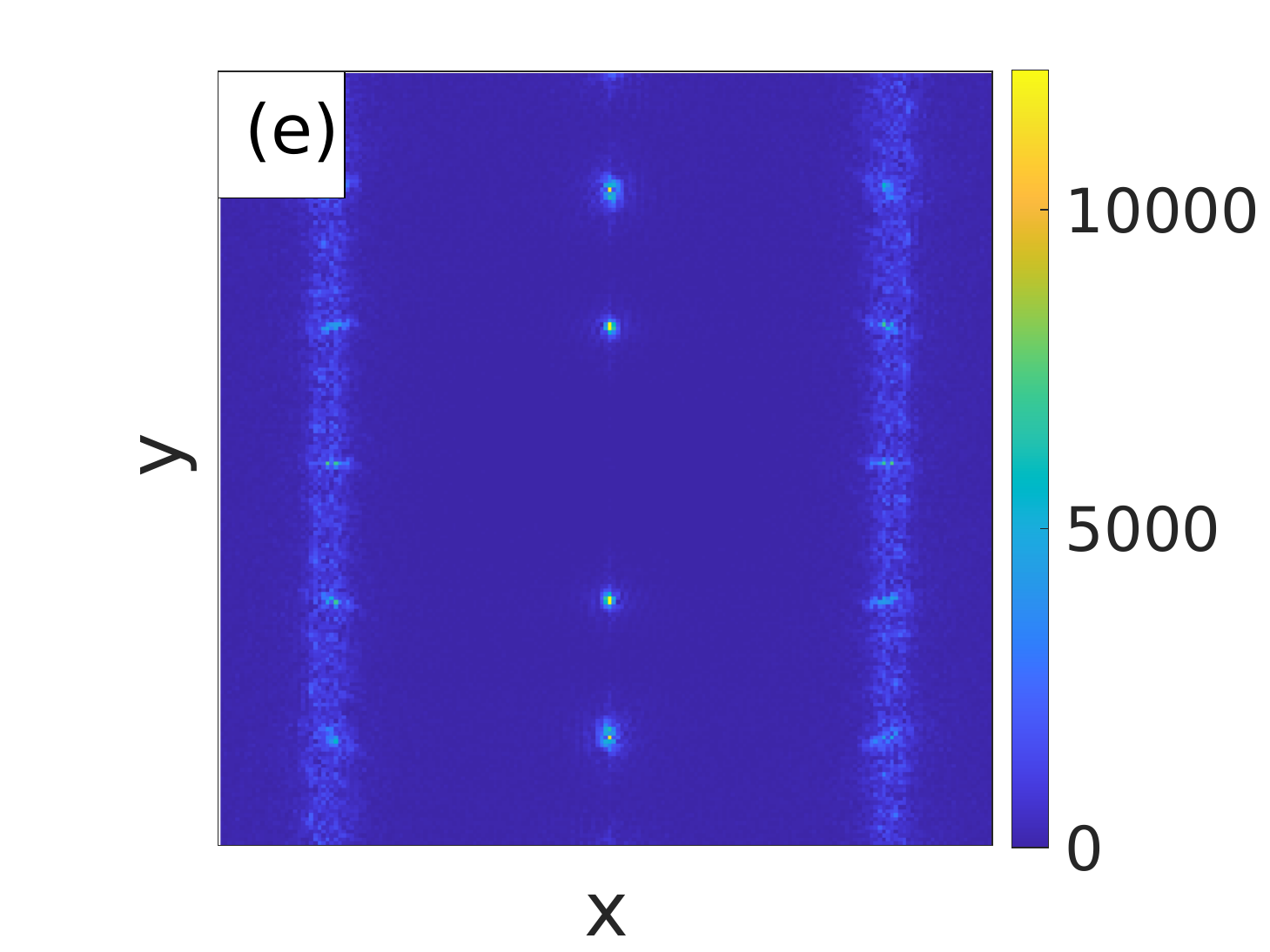}
  \end{minipage}\hfill\\
  \begin{minipage}{0.7\textwidth}
    \includegraphics[width=0.5\textwidth]{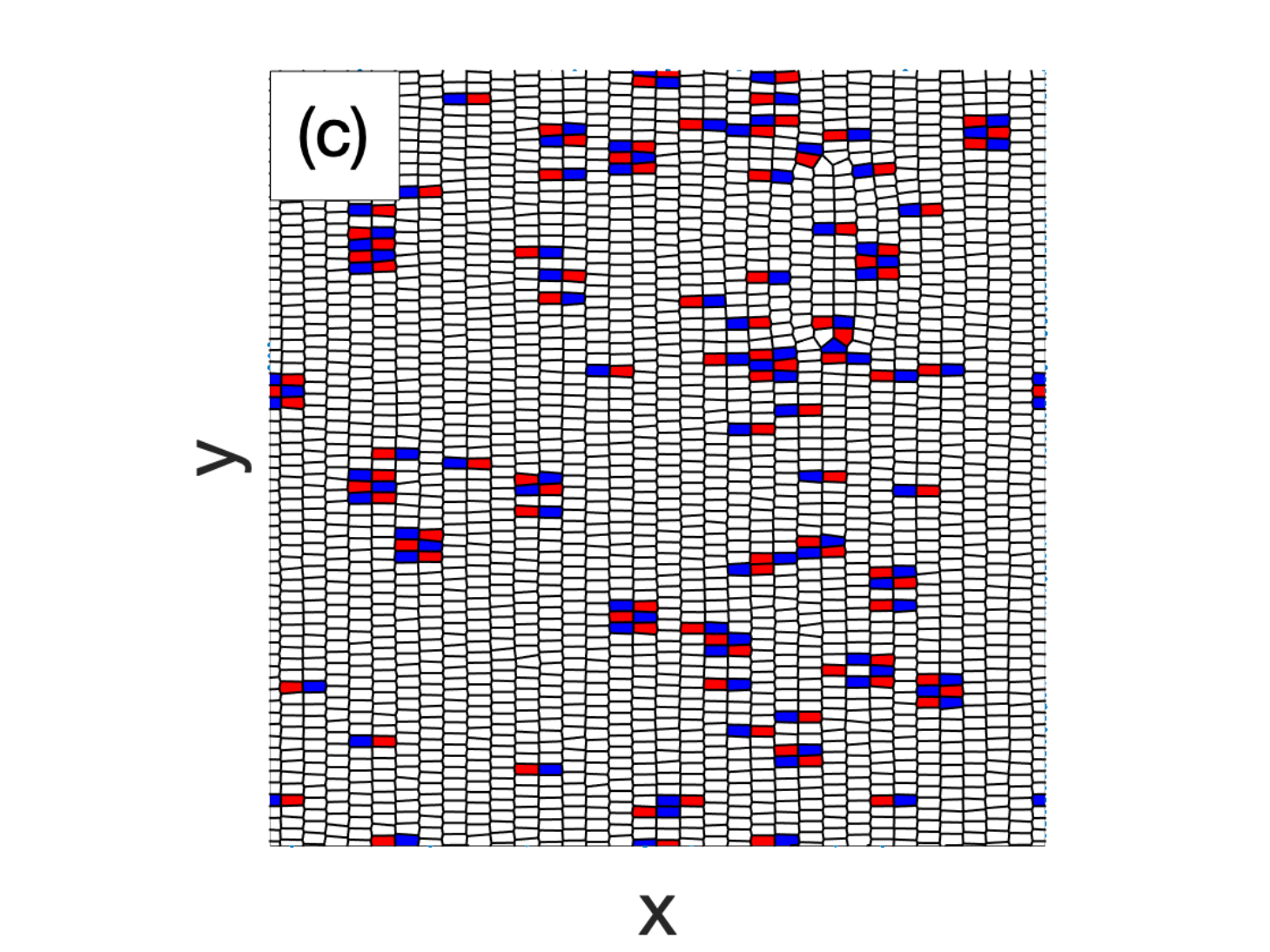}%
    \includegraphics[width=0.5\textwidth]{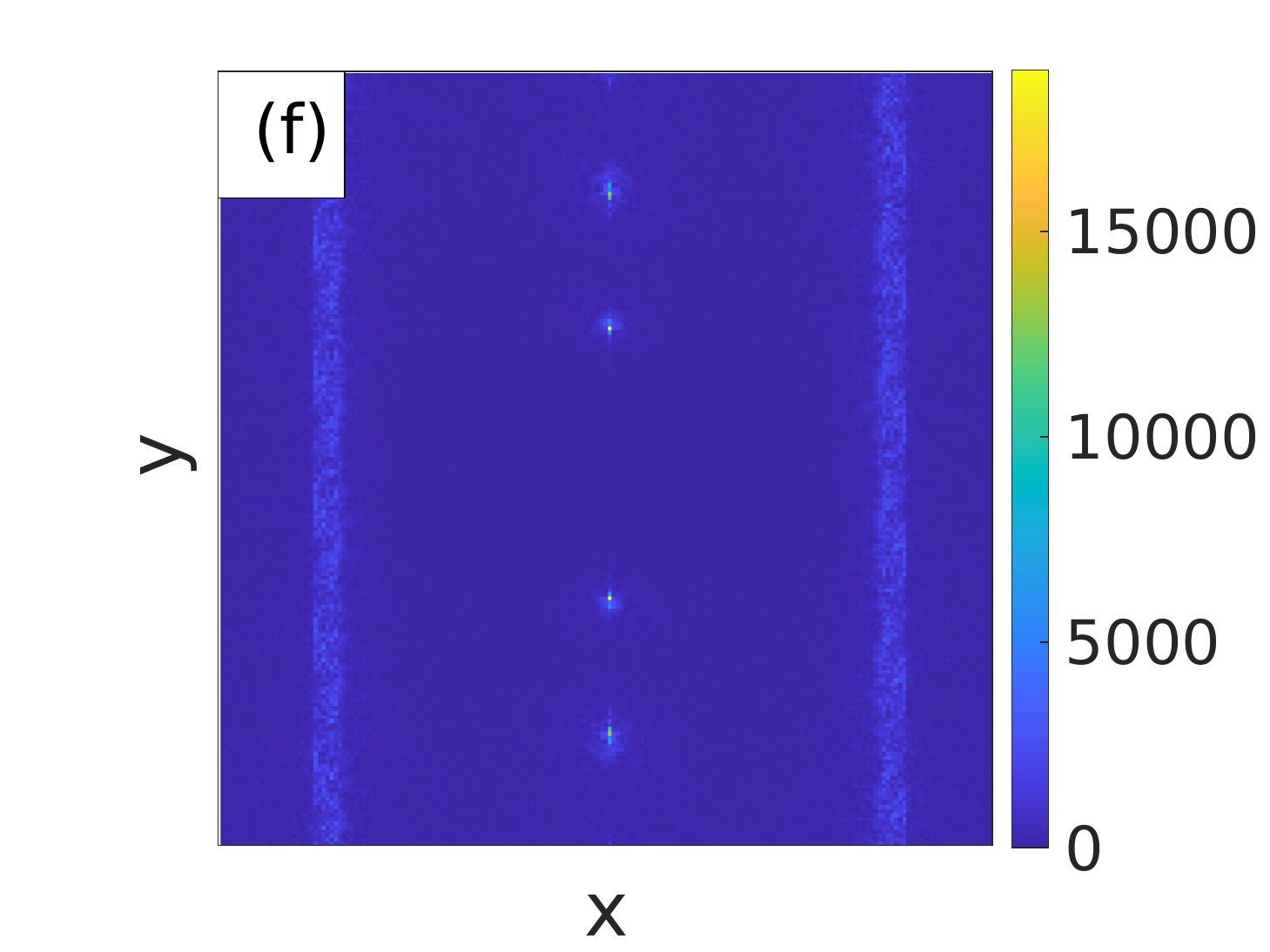}
  \end{minipage}\hfill
    \caption{
(a,b,c) Voronoi construction of a portion of the sample showing sixfold (white), fivefold (red), and sevenfold (blue) coordinated vortices for the system in Fig.~\ref{fig:2} with $\rho_p=0.48225$ and $K_2=0.55$. (d,e,f) The corresponding structure factor $S({\bf k})$.
(a,d) A pinned nematic phase at $F_{D} = 0.0$.
(b,e) A smectic structure forms under driving in the $x$-direction
at $F_D=1.5$.
(c,f) A similar smectic structure appears for driving in the $y$-direction
at $F_D=1.5$.}
    \label{fig:3}
\end{figure*}

In Fig.~\ref{fig:3}(a) we show
a Voronoi construction of a portion of the system
from Fig.~\ref{fig:2} with $K_2=0.55$
and $\rho_p=0.48225$ at $F_{D} = 0.0$,
where one-dimensional (1D) chains of vortices appear that are
aligned in the $y$-direction.
The corresponding structure factor $S(k)$ in
Fig~\ref{fig:3}(d)
contains a set of diffusive peaks aligned in the
$k_y$ direction along $k_x = 0$,
indicative of nematic ordering.
Here the vortices are spaced more closely in the $y$-direction than in the
$x$-direction
due to the anisotropy of the repulsive vortex-vortex force, which is
smaller along the $y$ direction, permitting the vortices to approach each
other more closely from this direction.
In Fig.~\ref{fig:3}(c,f) when a drive of $F_D=1.5$ is applied along
the $y$-direction,
there are only a small number
of dislocations present that are all aligned in the
$y$-direction.
The corresponding structure factor is still anisotropic but
has sharp peaks in the
$y$-direction indicative of a smectic phase.  
Figure~\ref{fig:3}(b,e) shows the same
system with a drive of $F_{D} = 1.5$ applied along the $x$-direction.
A similar moving smectic
appears that is perpendicular to the drive direction.

\begin{figure*}
  \centering
  \includegraphics[width=\textwidth]{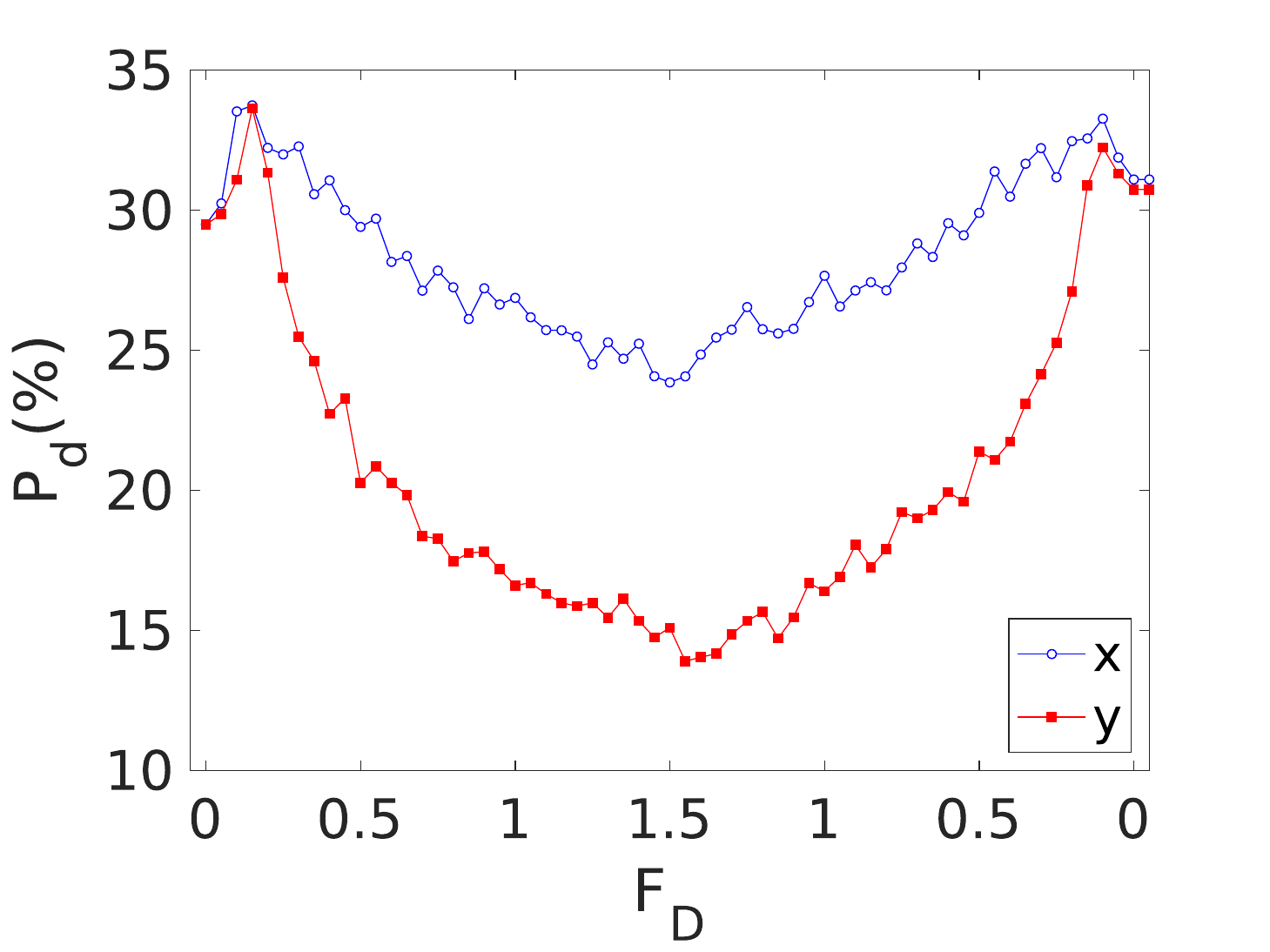}
    \caption{
$P_{d}$ versus $F_{D}$
in a system with $K_2 = 1.45$ and $\rho_p = 0.48225$ for $x$-direction driving (blue)
and $y$-direction driving (red).
The
system is more ordered for driving in the $y$-direction.
    }
    \label{fig:4}
\end{figure*}

In Fig.~\ref{fig:4} we plot $P_{d}$ versus $F_{D}$
for driving in the $x-$ and $y$-directions for
a sample with $\rho_{p} = 0.48225$ at 
a larger anisotropy of $K_{2} = 1.45$.
There is a clear difference in the defect density, with
$y$-direction driving producing much lower values of $P_d$ than
$x$-direction driving
over the range of drives we consider,
indicating that the
system is better ordered when the drive is aligned with the soft
anisotropy direction or the natural smectic orientation of the system.
For driving in the $x$-direction,
$P_{d}$ reaches a minimum
value of $P_d=0.24$,
while we find a lower minimum
of $P_d=0.14$ for driving in the $y$-direction.
As the drive is decreased from its maximum value of $F_D=1.5$,
the system starts to disorder again for
driving in both directions,
reaching
nearly identical values of $P_d \approx 0.32$ at the pinning transition.

\begin{figure*}
  \centering
  \begin{minipage}{0.6\textwidth}
    \hspace{0.15in}\includegraphics[width=0.4\textwidth]{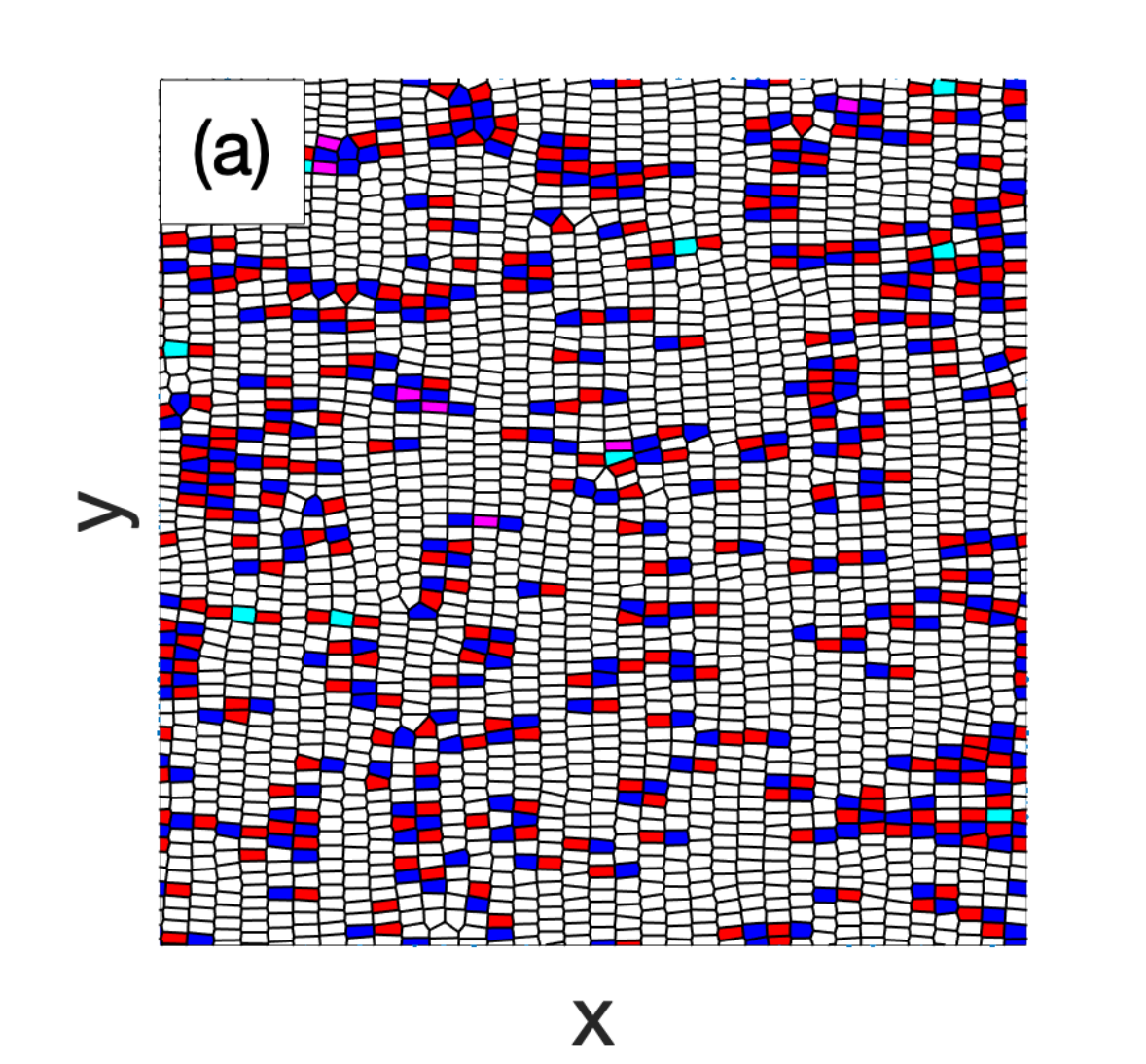}%
    \hspace{0.15in}\includegraphics[width=0.5\textwidth]{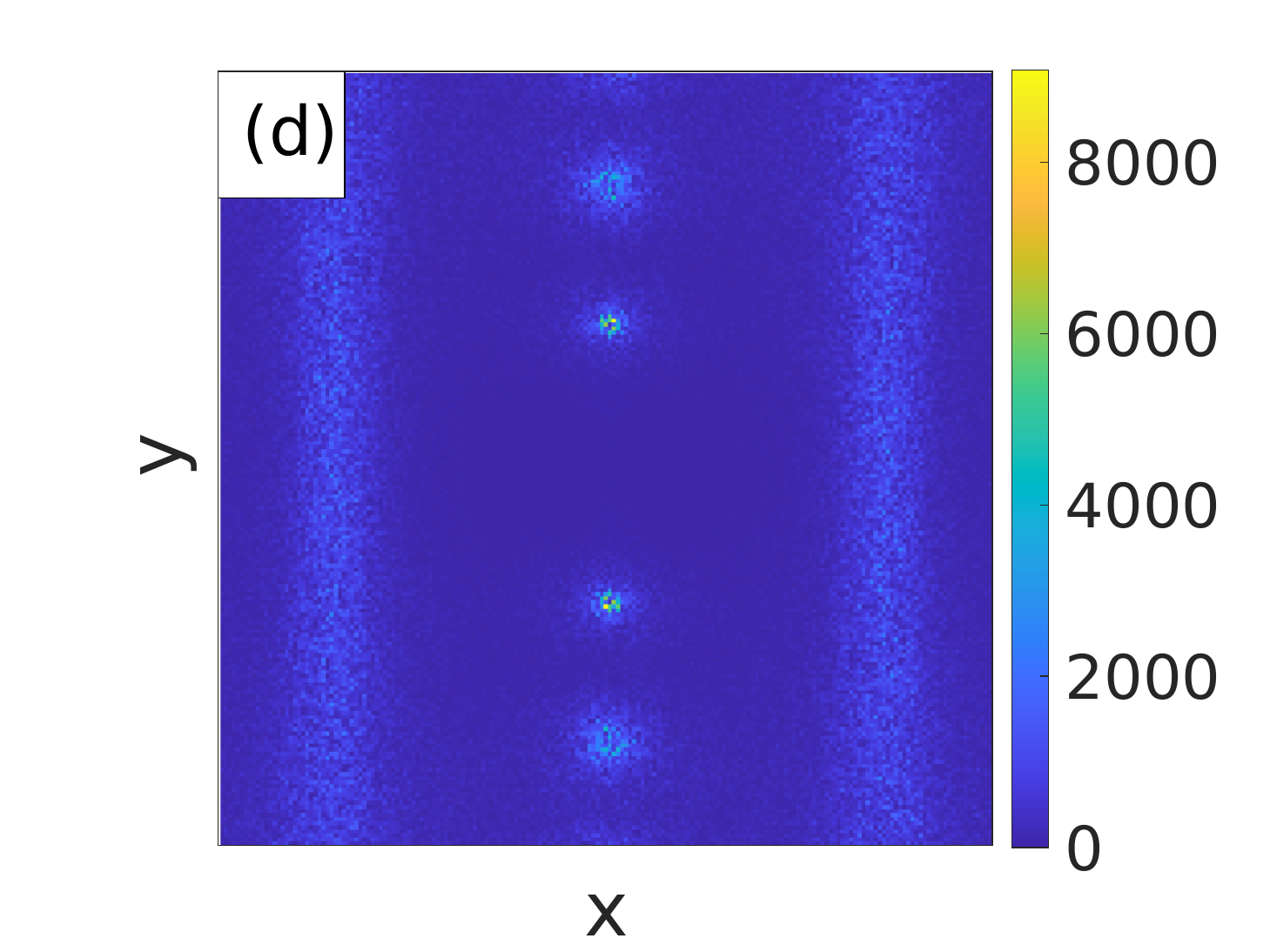}
  \end{minipage}\hfill\\
  \begin{minipage}{0.6\textwidth}
    \includegraphics[width=0.5\textwidth]{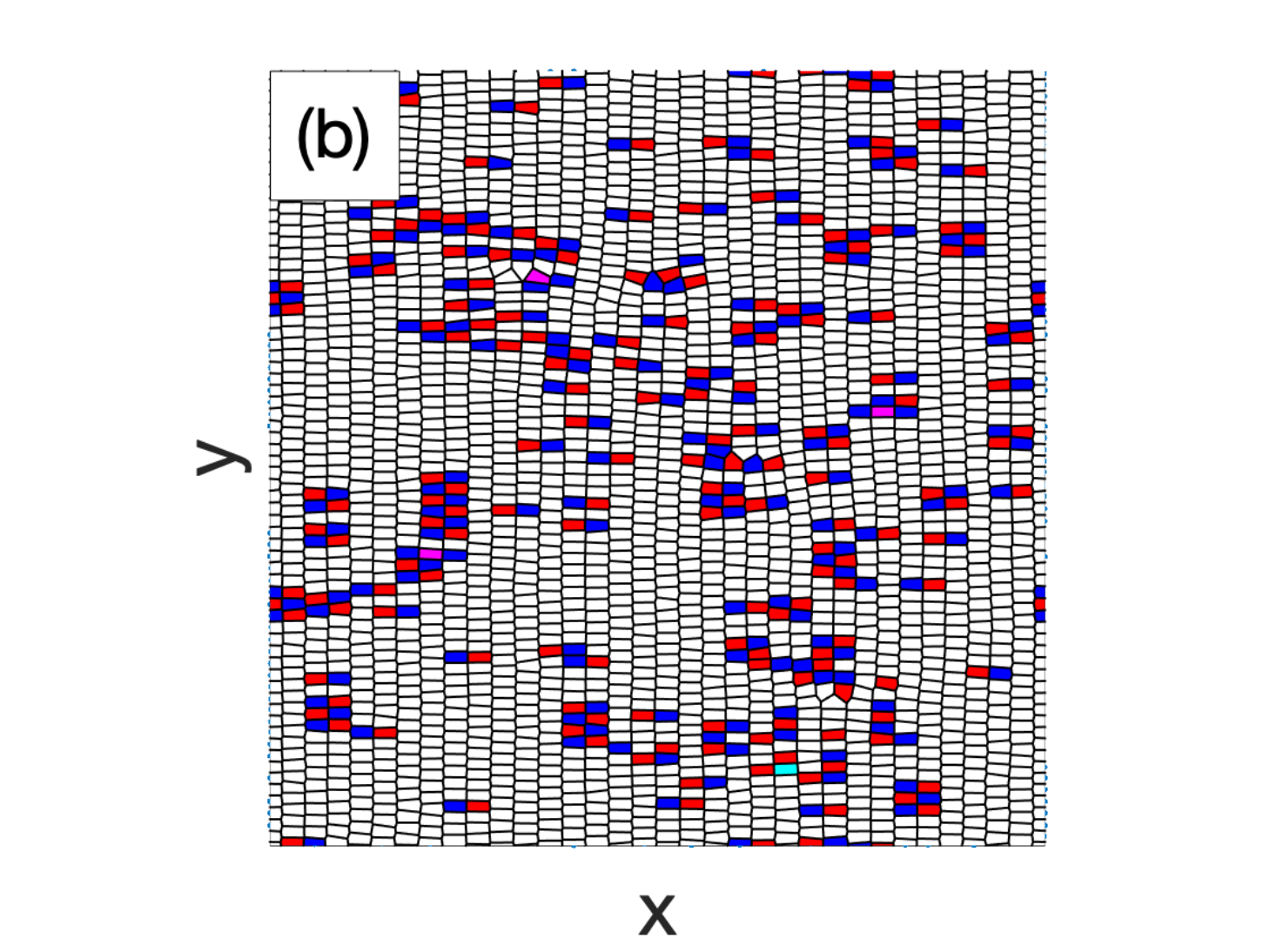}%
    \includegraphics[width=0.5\textwidth]{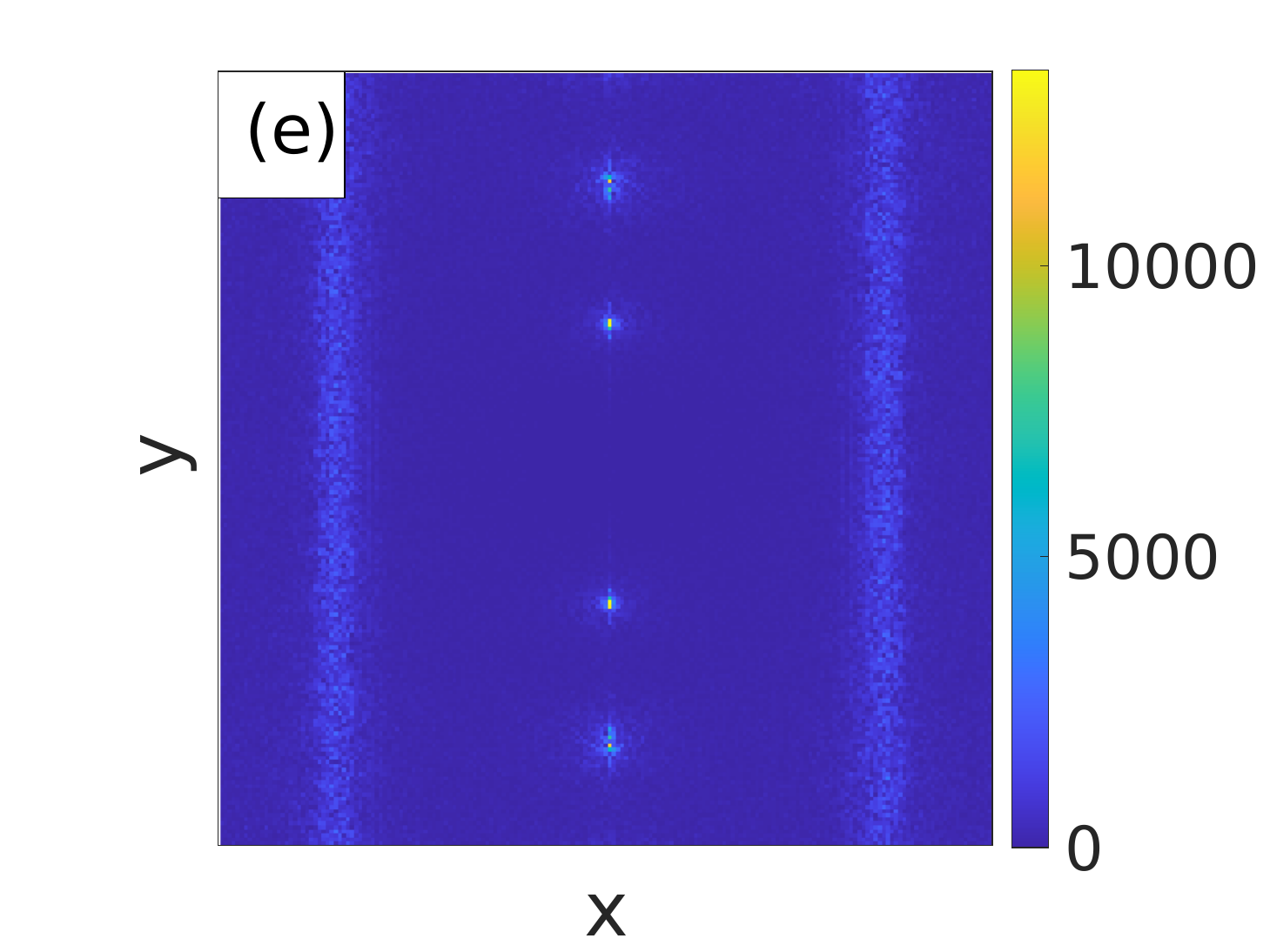}
  \end{minipage}\hfill\\
  \begin{minipage}{0.6\textwidth}
    \includegraphics[width=0.5\textwidth]{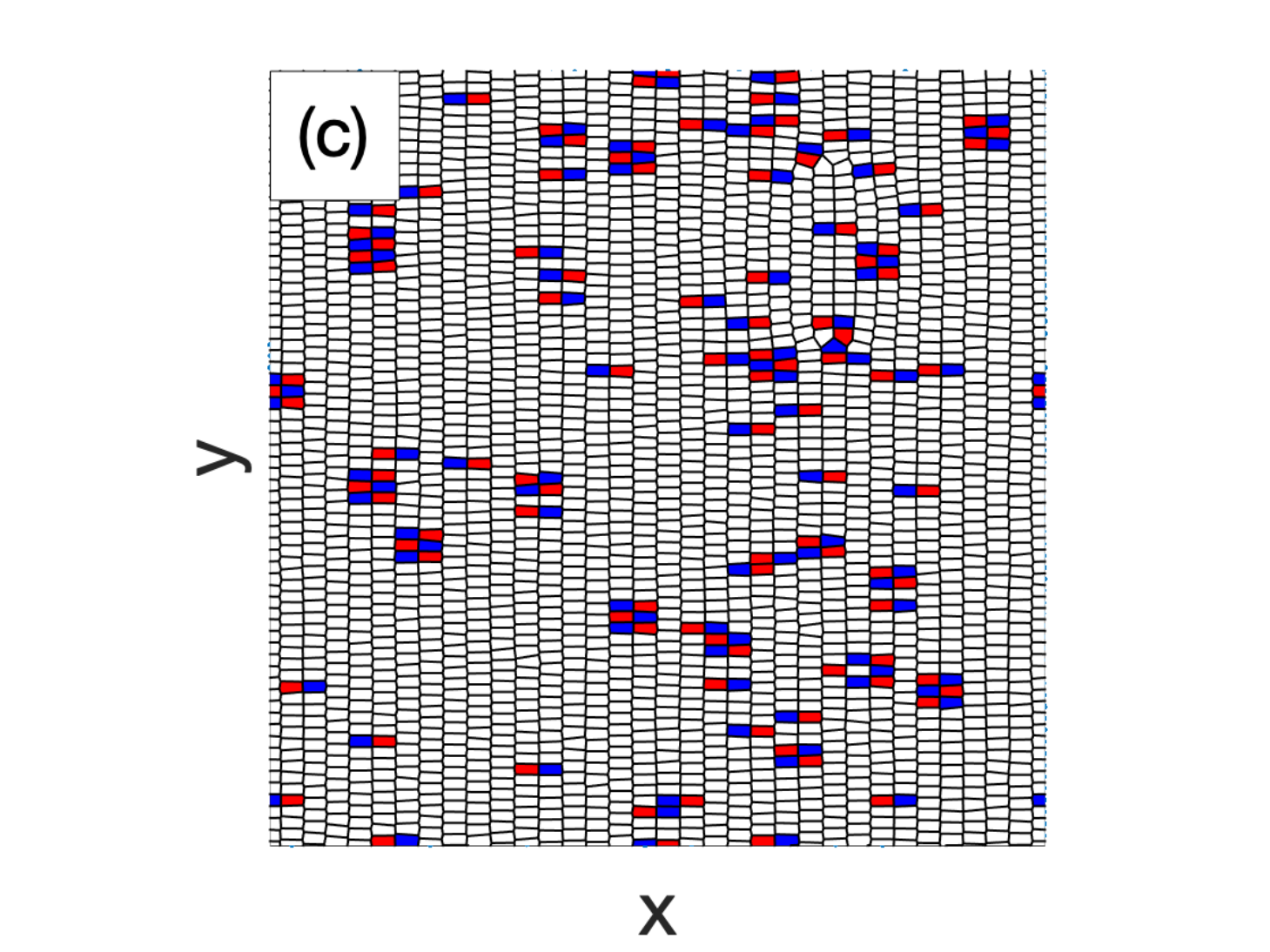}%
    \includegraphics[width=0.5\textwidth]{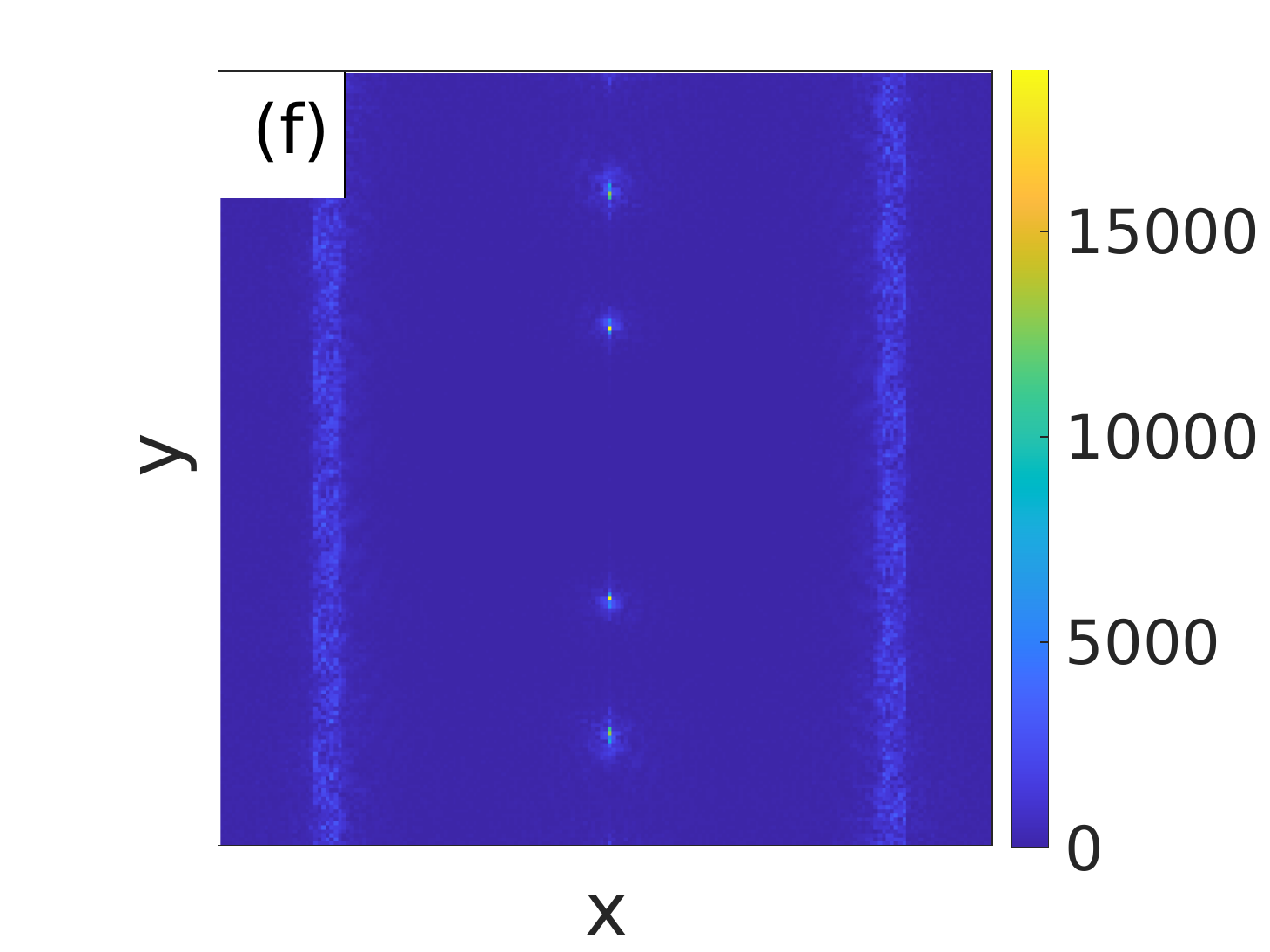}
  \end{minipage}\hfill
  \caption{
(a,b,c) Voronoi construction of a portion of the sample showing sixfold (white),
fivefold (red), and sevenfold (blue) coordinated vortices for the system
in Fig.~\ref{fig:4} with $\rho_p=0.48225$ and $K_2=1.45$. (d,e,f) The corresponding structure
factor $S(k)$.
(a,d) Nematic ordering at $F_D=0$.
(b,e) Driving in the $y$-direction at $F_D=0.75$.
(c,f) A smectic state forms for driving in the $y$-direction
at $F_D=1.5$.
  }
    \label{fig:5}
\end{figure*}

In Fig.~\ref{fig:5}(a,d) we show a Voronoi construction and structure factor
for the
system in Fig.~\ref{fig:4} with $K_2=1.45$
at $F_{D}=0$, where a pinned nematic phase appears. 
Figure~\ref{fig:5}(b,e) shows the same system
for driving in the $y$-direction with $F_{D} = 0.75$. 
The vortices are more ordered, as indicated by the sharper peaks
in $S(k)$,
and the system has formed a moving nematic.
In Fig.~\ref{fig:5}(c,f), for driving in the $y$-direction
at $F_{D} = 1.5$,
the number of defects has diminished
and the peaks in $S(k)$ are sharper.
The vortices exhibit smectic ordering and form a series of non-overlapping
1D chains.

\begin{figure*}
\begin{minipage}{\textwidth}
  \includegraphics[width=0.35\textwidth]{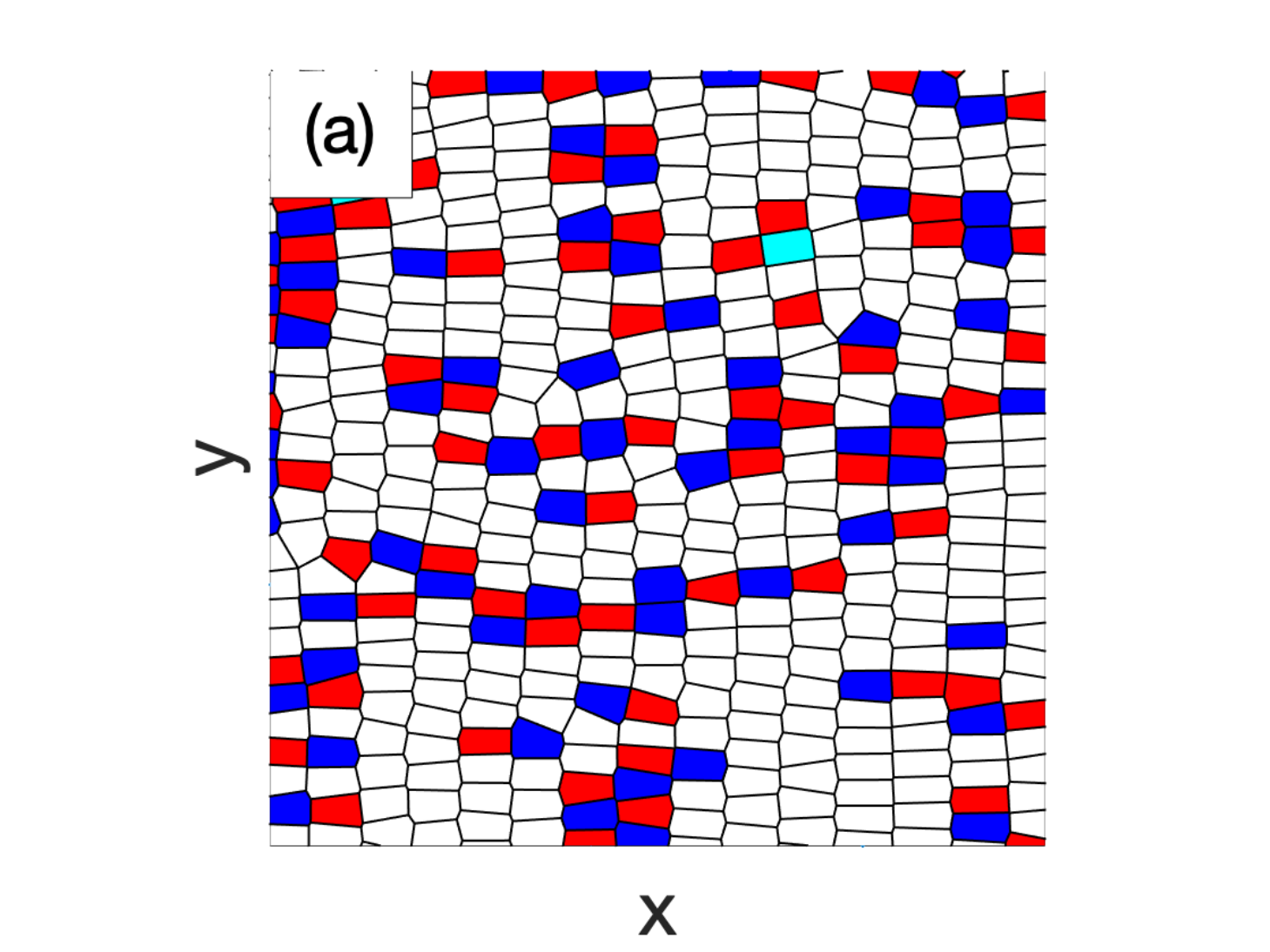}%
  \includegraphics[width=0.38\textwidth]{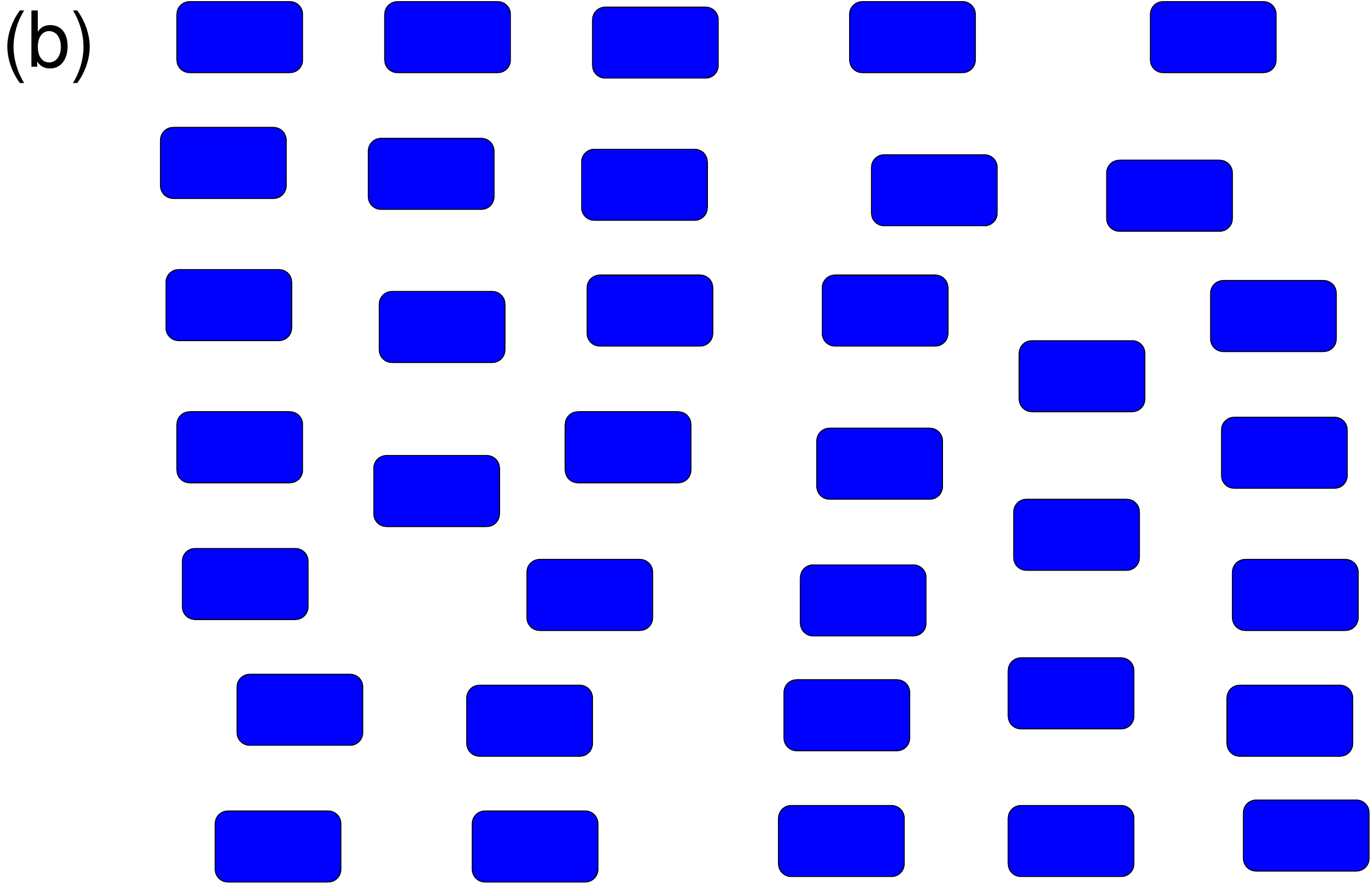}%
\end{minipage}\vspace{0.1in}
\begin{minipage}{\textwidth}
  \includegraphics[width=0.35\textwidth]{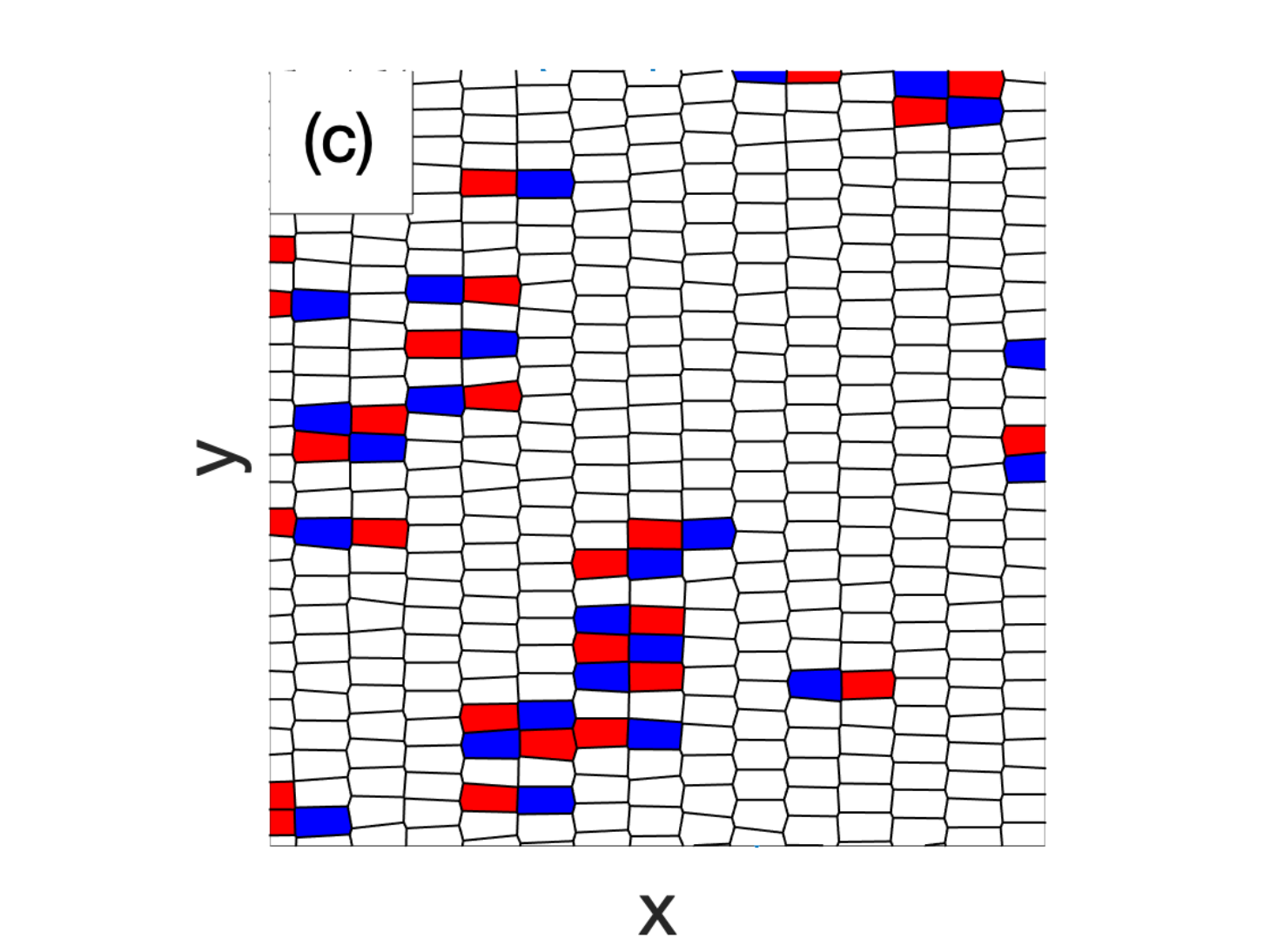}%
  \includegraphics[width=0.38\textwidth]{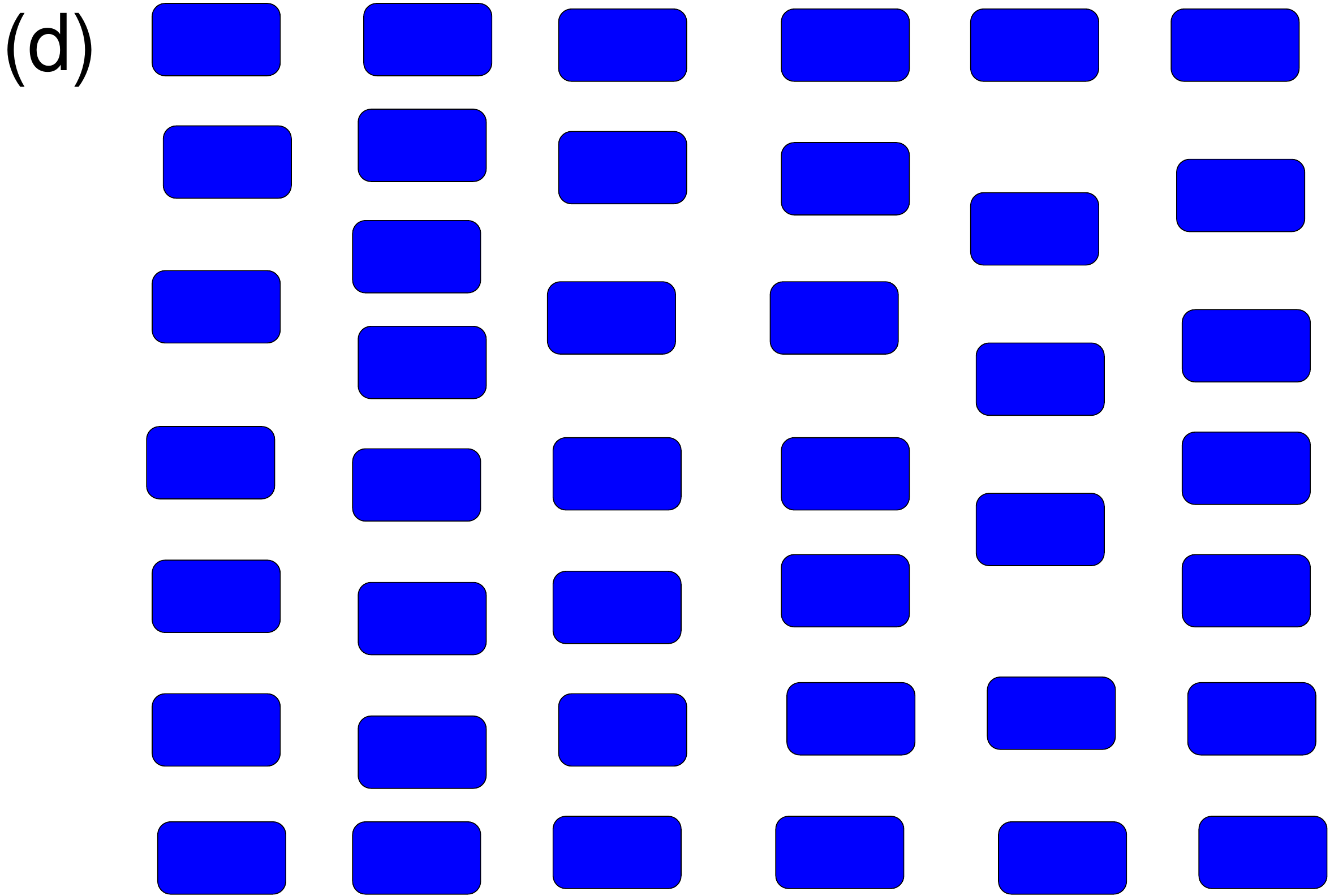}%
\end{minipage}
\caption{
(a) Exploded view of the Voronoi construction from Fig.~\ref{fig:5}(a)
for the sample with $\rho_p=0.48225$ and $K_2=1.45$ at $F_D=0$ showing 
the 1D chains of vortices that can break or merge. 
(b) A schematic of the vortex locations 
showing a nematic arrangement.
(c) Image of the vortex positions for the system in
Fig.~\ref{fig:5}(c,f) under $y$-direction driving with $F_{D} = 1.5$, where
the system forms 1D chains that do not cross.
(d) A schematic of the vortex locations
showing a smectic-A arrangement. }
    \label{fig:6}
\end{figure*}

In Fig.~\ref{fig:6}(a) we show a
smaller region of the Voronoi construction
from Fig.~\ref{fig:5}(a)
at $F_{D} = 0.0$ indicating that the vortices
form chains that can break or intertwine. 
Figure~\ref{fig:6}(b) shows a schematic of the 
vortices with the
anisotropic potentials forming a nematic structure.
Here the system forms chains that can end or begin inside the sample. 
In Fig.~\ref{fig:6}(c)
we show a small region of the Voronoi construction from
Fig.~\ref{fig:5}(e) 
for $y$-direction driving with
$F_{D} = 1.5$ where the system forms a smectic state and the
1D chains do not overlap.
Figure~\ref{fig:6}(d) shows a schematic of the
vortex structure in this state,
which is known as
smectic-A
in liquid crystal systems \cite{Azaroff80}.
This is similar to the phase proposed for
vortex liquid crystals with anisotropic potentials \cite{Carlson03}.
Here there are no breaks in the 1D chains.
Individual chains can contain different numbers of vortices,
producing
dislocations that are aligned in the $y$-direction. 
In the $K_2 = 1.45$ system,
driving in the $x$-direction produces
a set of phases very similar to those found for
driving in the $y$-direction,
but the nematic phase persists up to higher drives.  

\begin{figure*}
  \begin{minipage}{\textwidth}
    \includegraphics[width=0.5\textwidth]{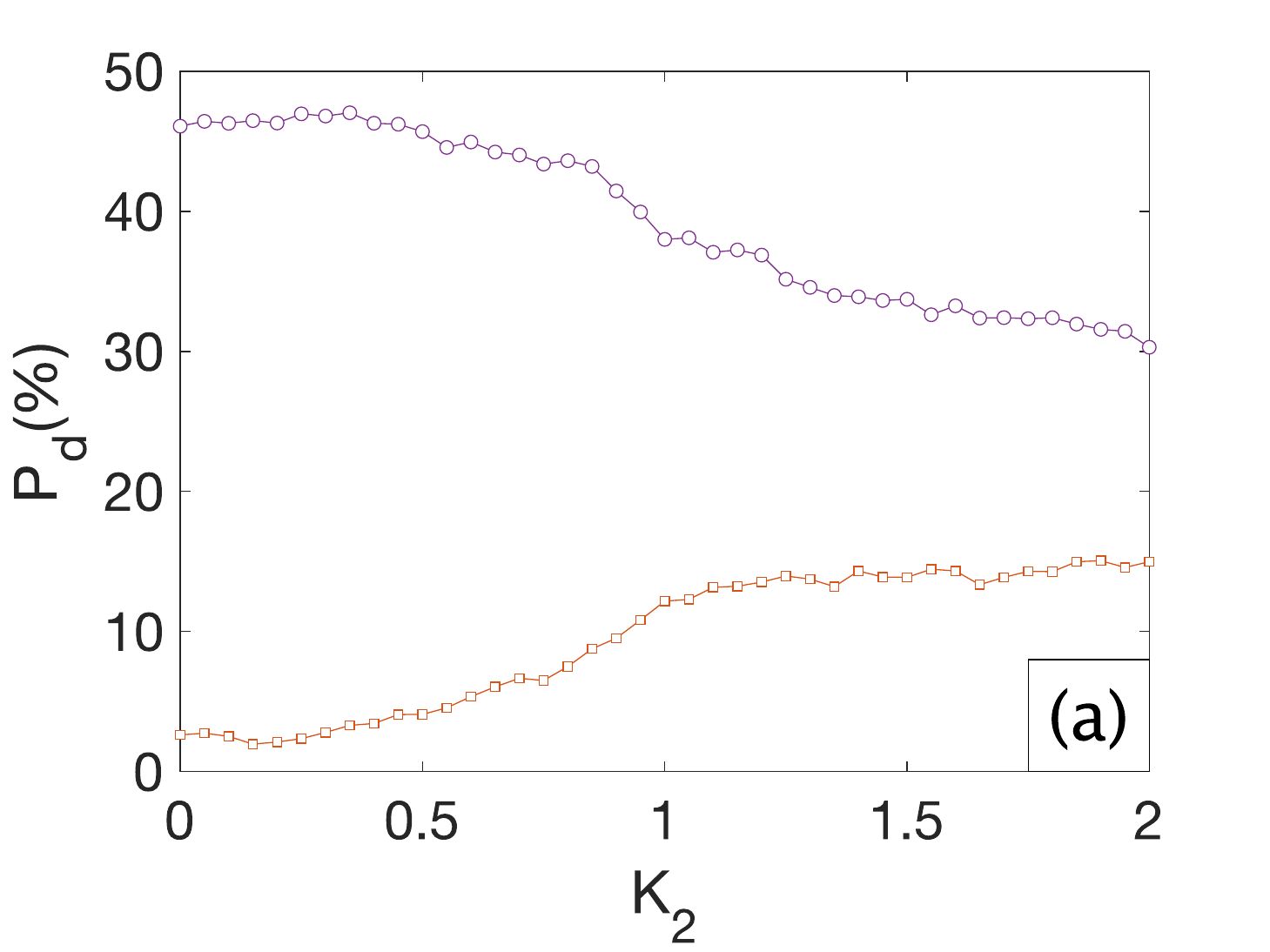}%
    \includegraphics[width=0.5\textwidth]{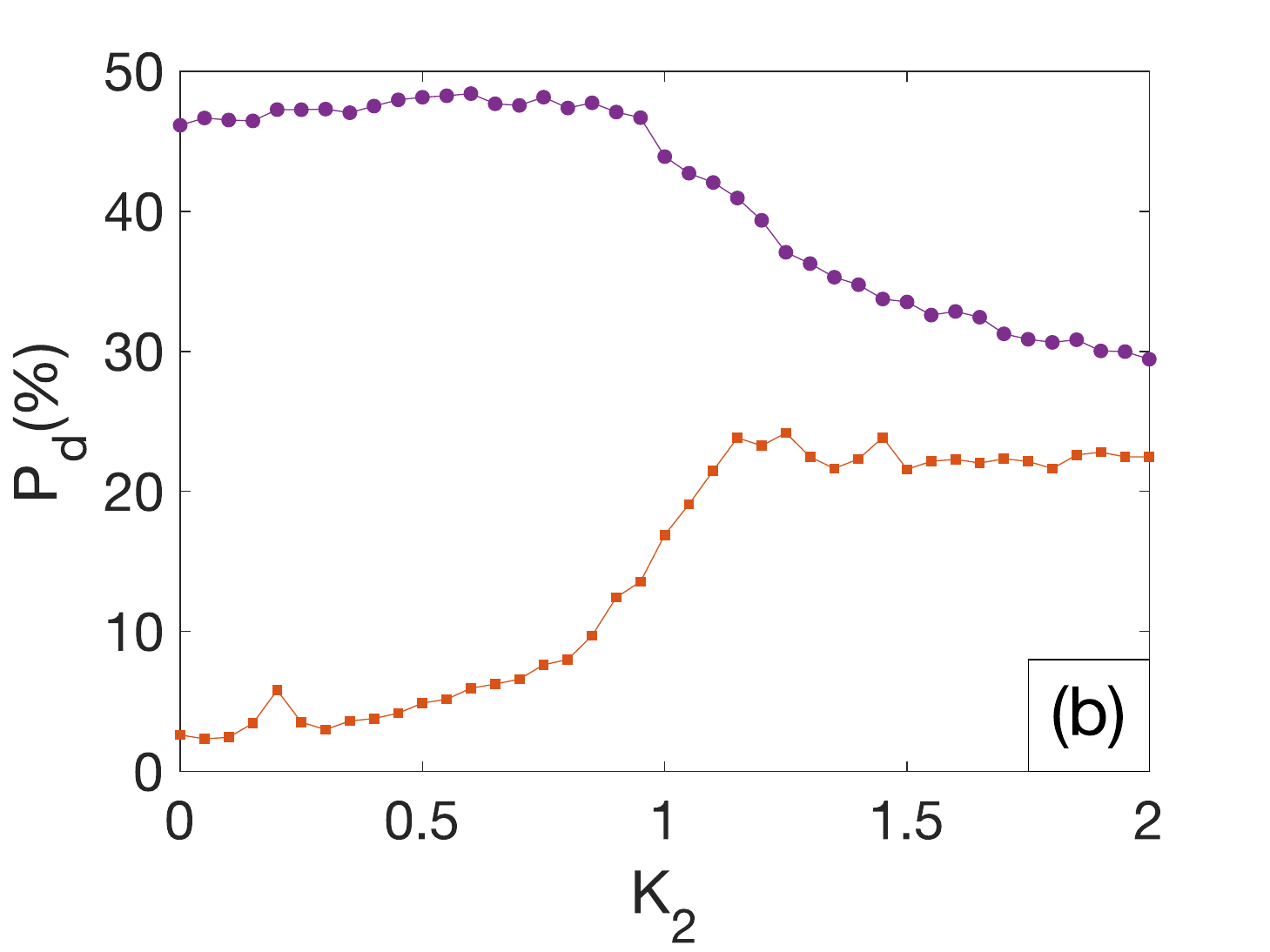}
  \end{minipage}
  \caption{
The minimum (red) and maximum (blue) values of $P_{d}$ versus
$K_{2}$
in samples with
$\rho_{p} = 0.48225$.
(a) Driving in the $y$-direction, and 
(b) in the $x$-direction.
}
  \label{fig:7}
\end{figure*}

We can also characterize the system by measuring the
maximum and minimum number of defects generated during the
drive cycle $F_D=0 \rightarrow 1.5 \rightarrow 0$ for varied $K_{2}$.
The maximum number of defects appear
at the depinning threshold, while the minimum number are present
at the highest drive of $F_{D}=1.5$. 
In Fig.~\ref{fig:7}(a) we plot the minimum and maximum
values of $P_d$ versus $K_2$ for
driving in the $y$-direction in samples
with $\rho_{p} = 0.48225$.
For low $K_2 < 0.8$, at depinning
there is a nematic state with $P_d\approx 0.45$.
The vortices
order into a moving smectic phase at higher drives
with the defect density reaching minimum values of
$P_d=0.03$ to $P_d=0.07$.
For
$K_{2} > 0.9$, the system is less defected at the depinning
transition but contains
more defects in the driven reordered states.
As the anisotropy increases, it becomes more difficult to destroy
the chains of vortices at depinning, giving a lower density of defects
at the depinning transition;
however, it becomes easier for the chains to slide past one another at
higher drives,
creating a larger number of more persistent
dislocations in the driven phase. 
In Fig.~\ref{fig:7}(b) we show the minimum and maximum values of $P_d$
versus $K_2$ 
for the same system under driving in the $x$-direction.
Near $K_2=0.2$ there is a peak
in the minimum number of defects corresponding to the critical
anisotropy at which the smectic undergoes
a transition from alignment in the $x$-direction to
alignment in the $y$-direction.
For $K_{2} > 0.2$ the system forms a moving smectic aligned in
the $y$-direction, while 
when $K_{2} > 0.9$, a moving nematic appears at higher drives. 

\begin{figure*}
\begin{minipage}{0.9\textwidth}
    \includegraphics[width=0.5\textwidth]{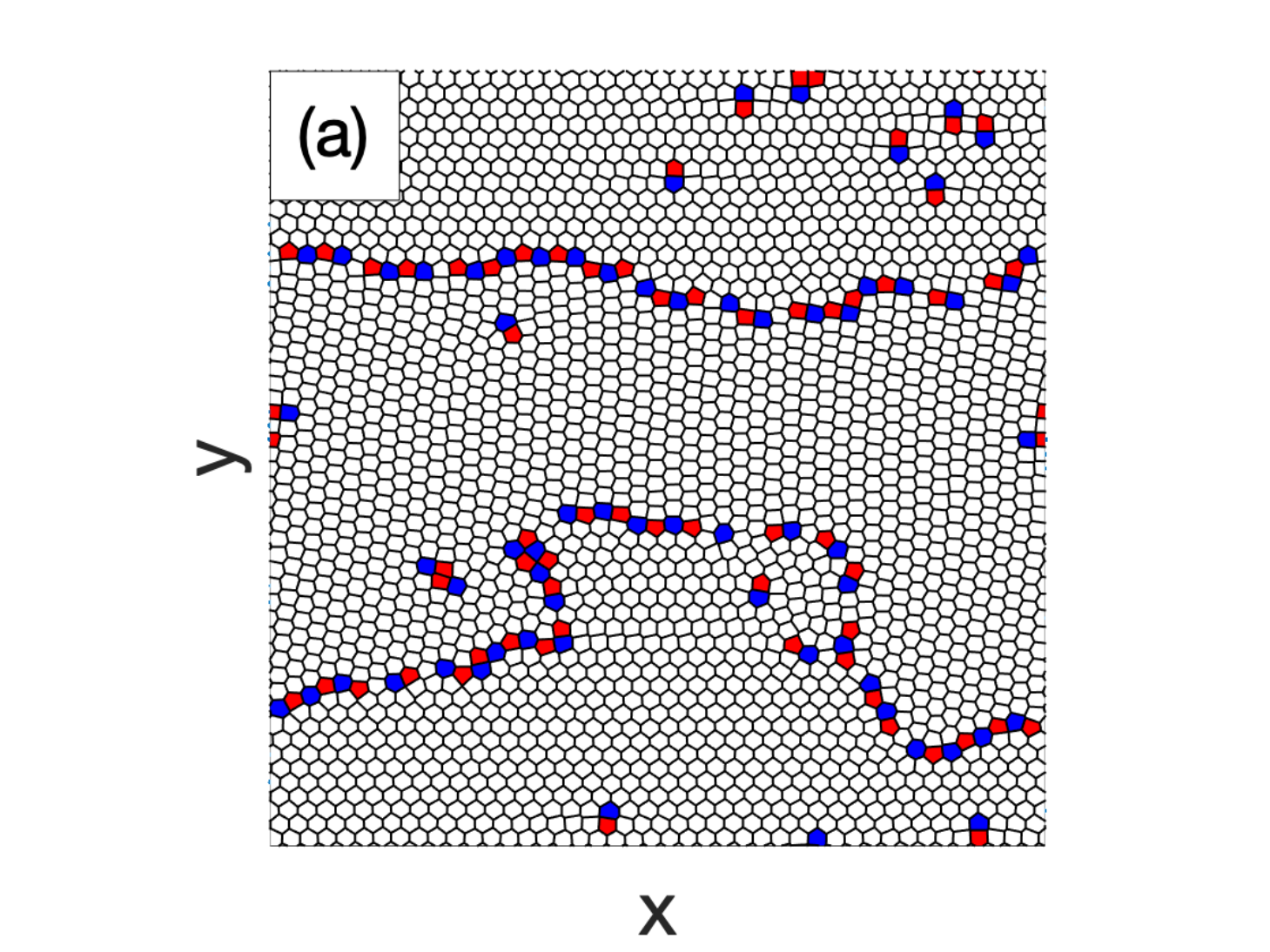}%
    \includegraphics[width=0.5\textwidth]{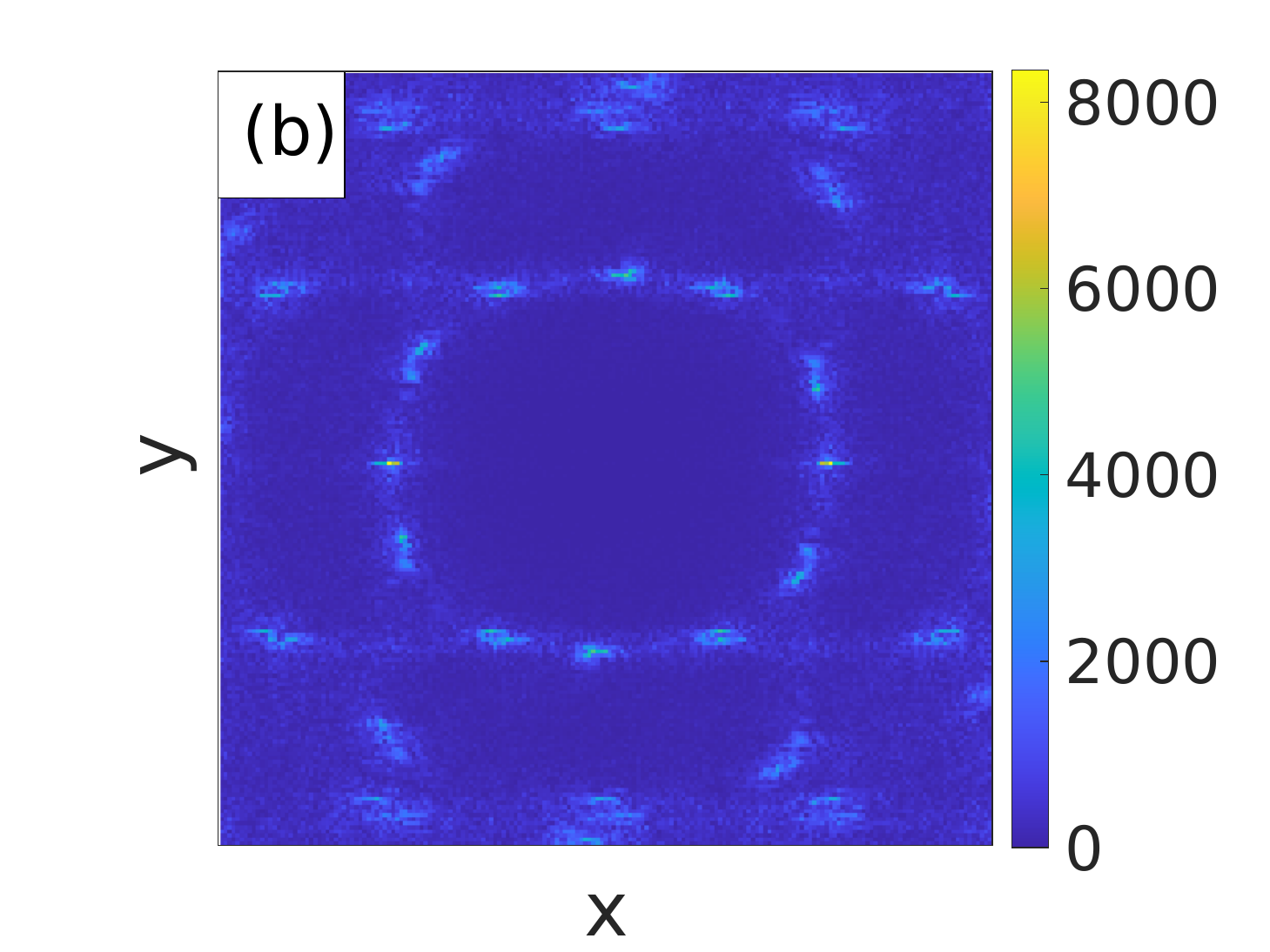}
\end{minipage}
    \caption{
(a)  Voronoi construction of a portion of the sample showing sixfold (white),
fivefold (red), and sevenfold (blue) coordinated vortices for the      
system in Fig.~\ref{fig:7}(b) with $\rho_p=0.48225$ and $K_2=0.2$ under
driving in the $x$-direction at $F_D=1.5$.
(b) The corresponding $S(k)$
shows a partial smectic alignment in the $x$-direction.
%\insert{[Can you add circles or arrows to indicate the smectc peaks in (b).]}}
 }
    \label{fig:8}
\end{figure*}

In Fig.~\ref{fig:8} we show the Voronoi construction and structure factor
for the system in Fig.~\ref{fig:7}(b) with $K_2=0.2$
for driving in the $x$-direction
at $F_{D} = 1.5$.
Here, the system does not form a nematic or
smectic aligned in the $y$-direction, but
instead adopts a polycrystalline ordering with a partial alignment in the
$x$-direction.
The ordering is more clearly visible in the
structure factor,
where
two prominent peaks are aligned with $k_x$.
For even smaller values of $K_{2}$,
the system exhibits a strong smectic alignment along the
$x$-direction for driving 
in the $x$-direction. 

\begin{figure}
    \centering
    \includegraphics[width=0.5\textwidth]{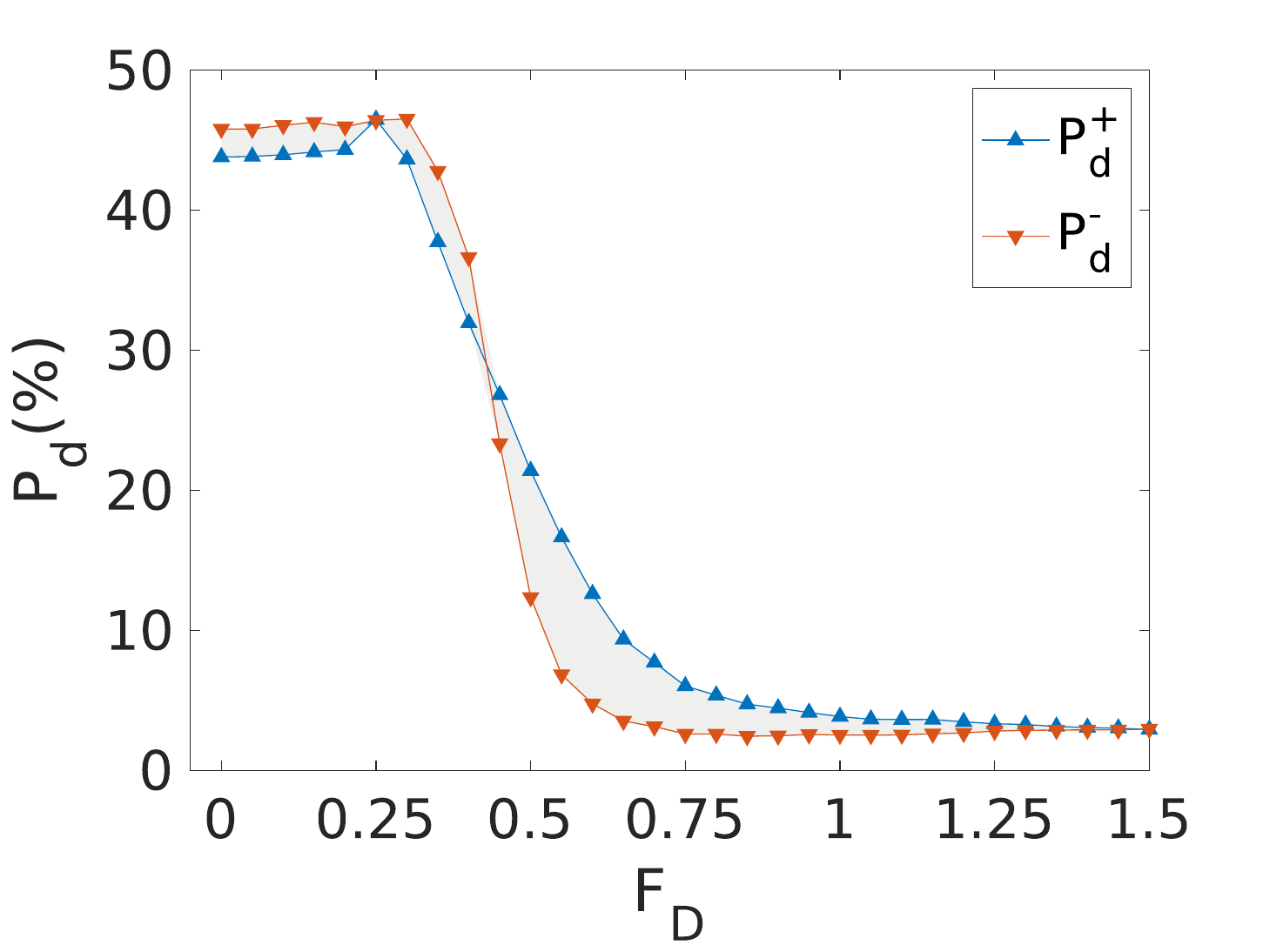}
\caption{An example of the curves used to construct the total hysteresis
measure $H$ defined in the text. $P_d$ versus $F_D$ is shown both for increasing current
($P_d^+$, blue) and
decreasing current ($P_d^-$, orange)
in a system with $K_2=0.1$, $\rho_p=0.48225$, and $x$-direction driving.
}
    \label{fig:hystoverlay}
\end{figure}

\begin{figure}
\begin{minipage}{\textwidth}
  \includegraphics[width=0.5\textwidth]{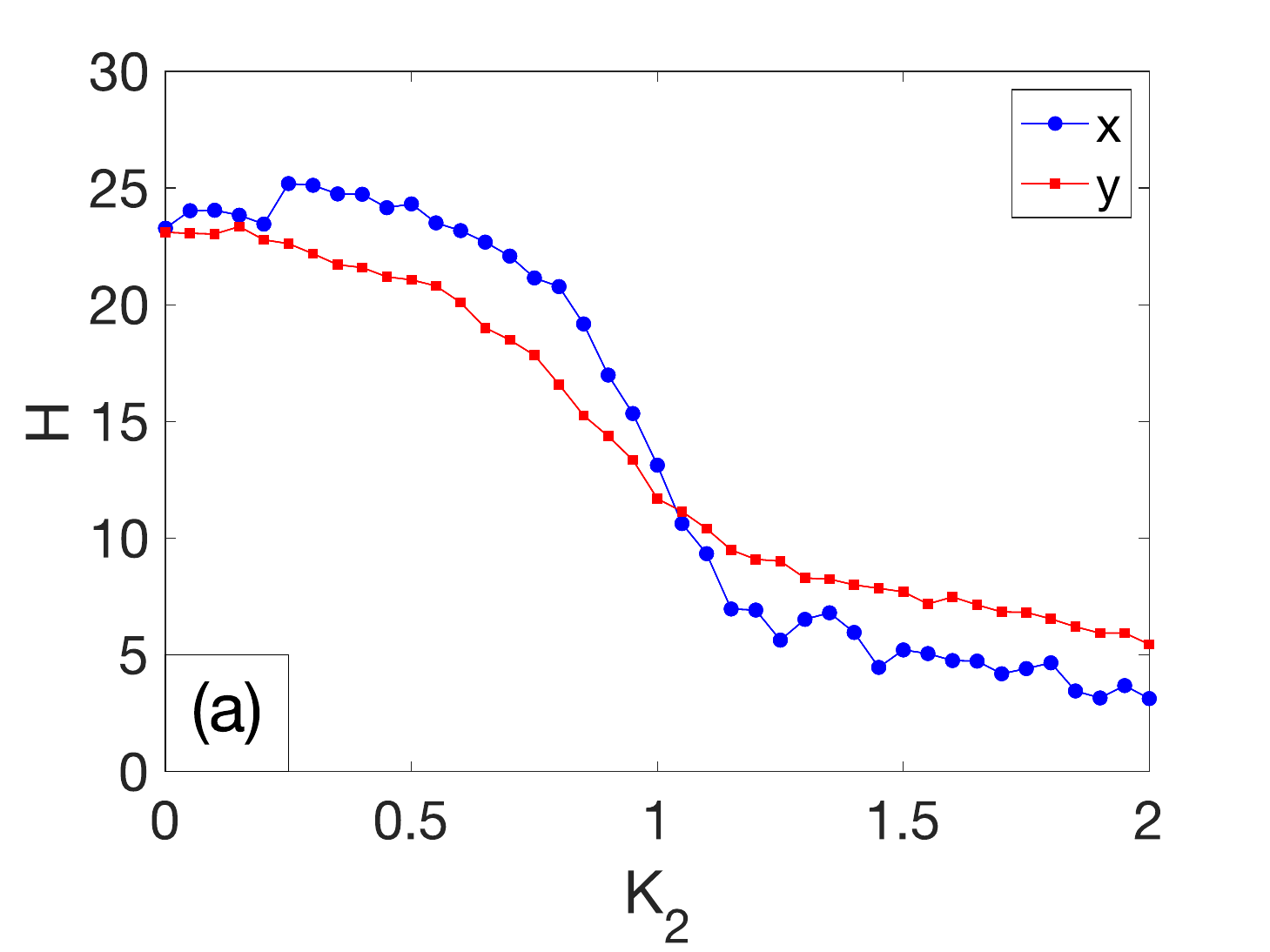}%
  \includegraphics[width=0.5\textwidth]{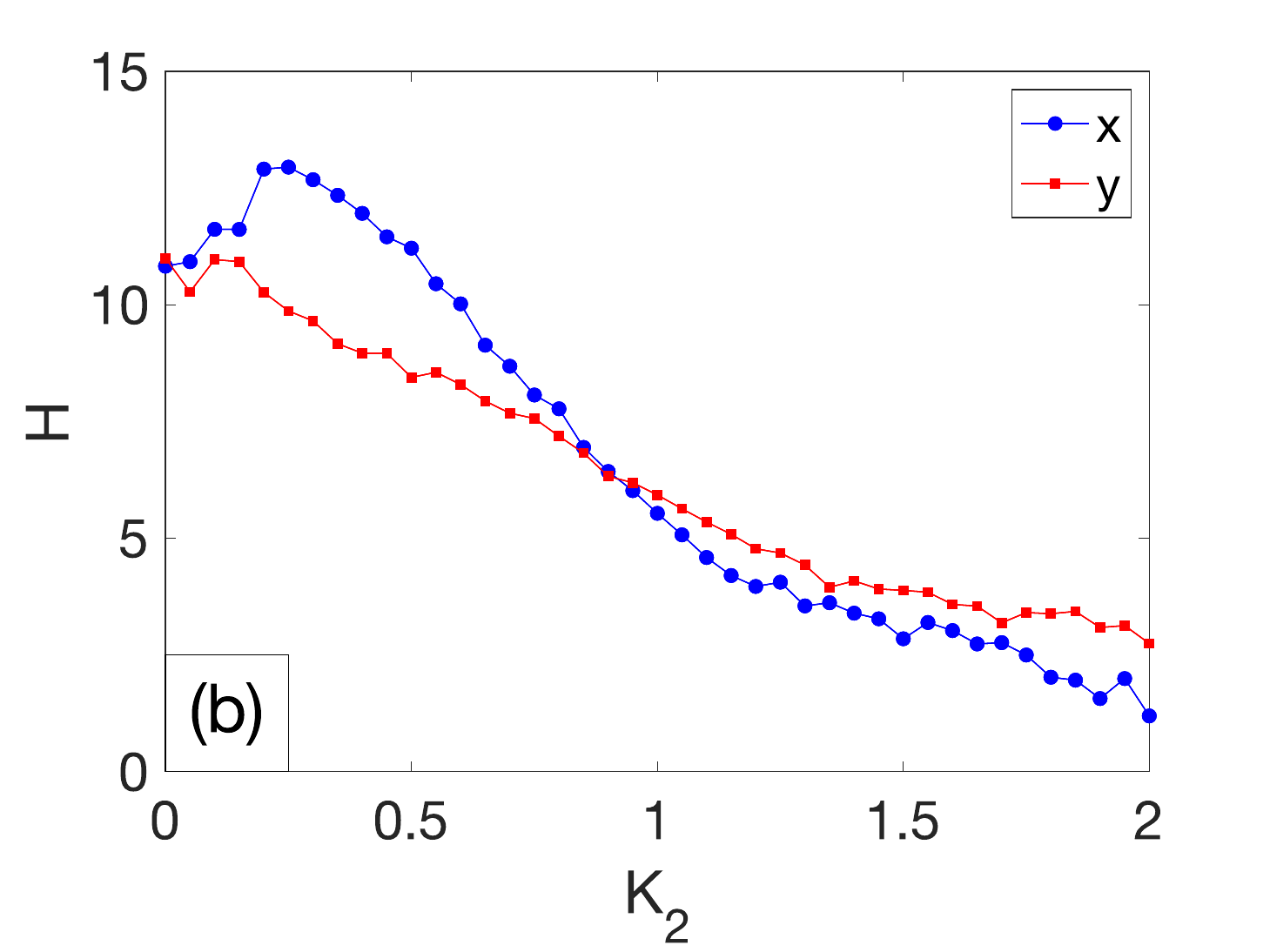}
\end{minipage}
\caption{
(a) The hysteresis $H$ versus $K_2$
  in a system with $\rho_{p} = 0.48225$.
  The blue curve is for driving in the $x$-direction and the red curve is
  for driving in the $y$-direction.
(b) The same for a system with $\rho_{p} = 0.09645$. }
    \label{fig:9}
\end{figure}

We can also characterize the system by measuring the 
total hysteresis in the form of
the sum 
of the absolute differences in $P_d$
for a given drive $F_D$, where we compare the value $P_d^+$ for
increasing current with $P_d^-$ for decreasing current.
The total hysteresis is obtained by numerically integrating the
absolute difference between the two curves,
$H=\int_{F_D=0}^{1.5} |P_d^+(F_D)-P_d^-(F_D)|$,
as indicated by the shaded area in
Fig.~\ref{fig:hystoverlay}.
We find that $H$
is largest
at lower values of the anisotropy, since in these systems
the number of defects varies over a greater range and
a nearly perfect lattice appears at high driving
that remains more robust against pinning forces
as the current is decreased.
In Fig.~\ref{fig:9}(a) we plot
the total hysteresis $H$ versus $K_2$ for a system with $\rho_{p} = 0.48225$
under driving in the $x$- and $y$-directions.
For driving in the $x$-direction, there is a peak
in $H$ near $K_{2} = 0.2$ corresponding to the smectic
$x$-direction to $y$-direction realignment
transition.
When $K_{2} < 1.0$,
the hysteresis is largest for driving in
the $x$-direction, but for larger $K_2$, we find
the largest hysteresis
for driving in the $y$-direction.
We observe similar behavior as we vary the pinning site density
$\rho_p$, with the
overall magnitude of $H$ gradually decreasing with decreasing $\rho_p$.
This is seen in Fig.~\ref{fig:9}(b), which shows $H$ versus $K_2$ at a
lower pinning density of $\rho_p=0.09645$,
where the total hysteresis is lower but the same trends appear.

\begin{figure}
\begin{minipage}{\textwidth}
\includegraphics[width=0.5\textwidth]{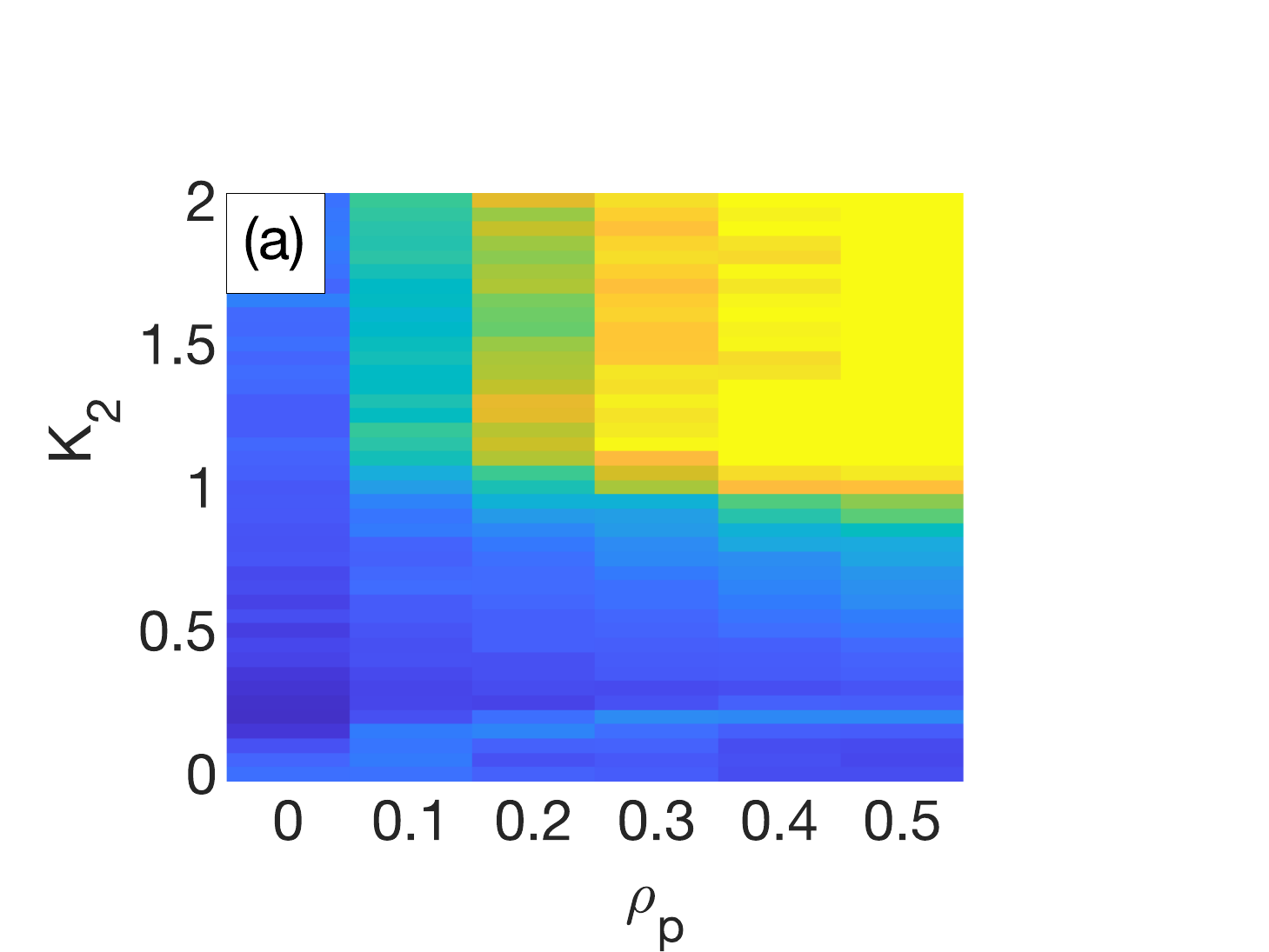}%
\includegraphics[width=0.5\textwidth]{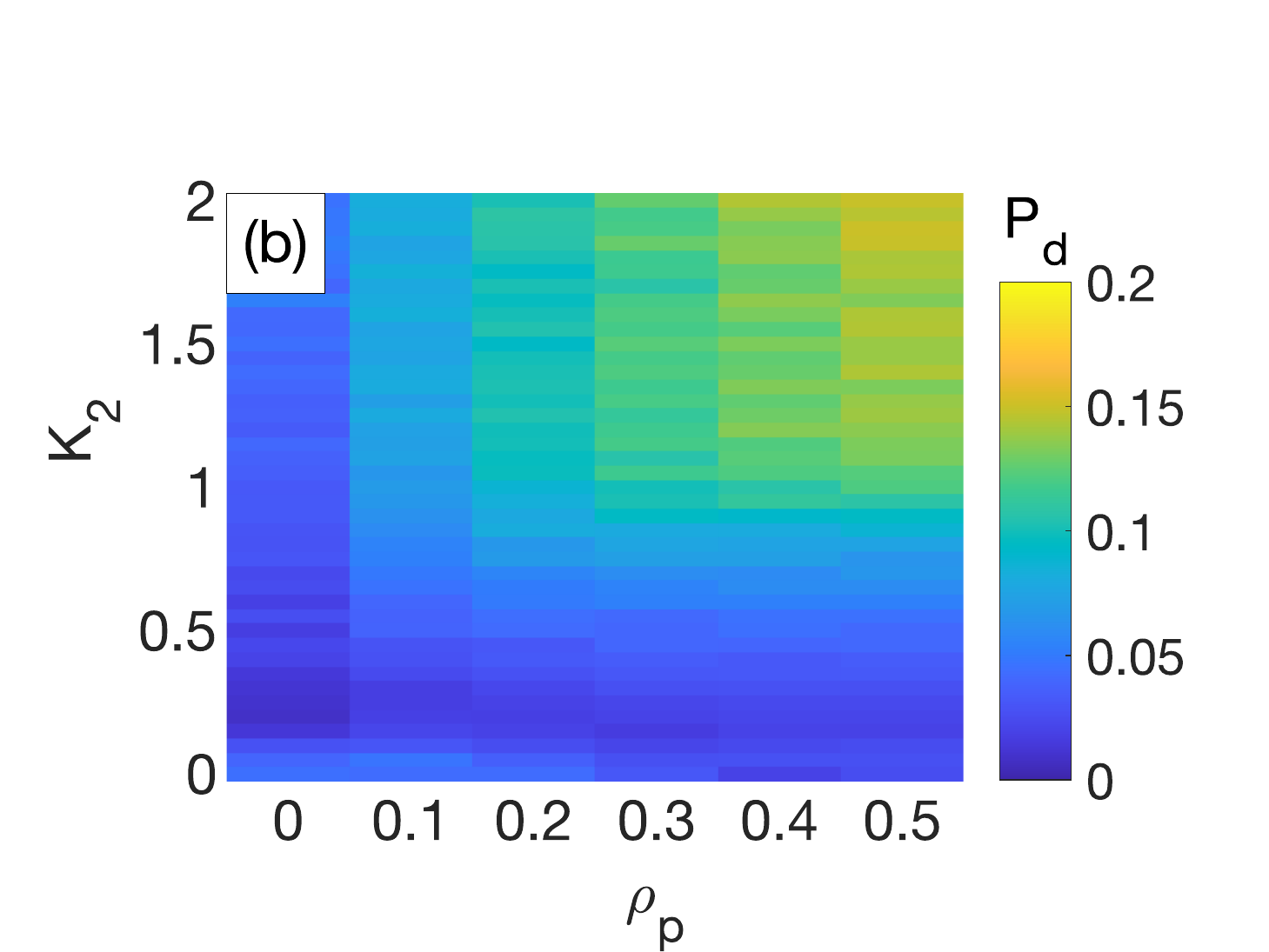}
\end{minipage}
\caption{
Heat map of the minimum number of topological defects $P_d$
at $F_d=1.5$ as a function
of $K_{2}$ and $\rho_{p}$.
(a) Driving in the $x$-direction. The light blue line in the lower region of
the plot indicates the transition from a smectic or nematic state aligned 
in the $x$-direction to a smectic or nematic state aligned in the $y$-direction.
(b) Driving in the $y$-direction.
%		\insert{[Add arrow or other indication of the transition feature in (a) - it's easy to miss.]}}
}
\label{fig:10}
\end{figure}

We have also considered the effect of changing the pinning density
over the range $\rho_p=0$ to $\rho_p=0.48225$.
In 
Fig.~\ref{fig:10}(a) we plot a heat map of the minimum value of $P_d$
({\it i.e.}, for $F_D=1.5$) as a function of
$K_{2}$ versus $\rho_{p}$
for driving in the $x$-direction.
The light blue line at $K_2 \approx 0.2$ indicates the
switching
of the smectic from
the $x$-direction to the $y$-direction alignment.
As $\rho_{p}$ increases, the critical value of
$K_{2}$ at which this transition occurs increases.
For $K_{2} > 1.0$ and $\rho_{p} > 0.1$, the amount of disorder
in the system increases dramatically and
a nematic structure is present,
while
in other regions of the parameter space,
we find a smectic state aligned in either the $x$- or $y$-directions.
Figure~\ref{fig:10}(b) shows the heat map of the minimum value of $P_d$
for driving in the $y$-direction.
In this case, the smectic structures formed by the vortices are always aligned
in the $y$-direction.
For $K_{2} > 1.0$ and $\rho_{p} > 0.19$,
there is an increasing number of defects in the moving phase,
but generally the system is in a smectic state.

%%% SUMMARY
\section{Summary}
We have examined the driven dynamics of vortices
with two-fold anisotropic interaction potentials
driven over quenched disorder.
In general, the pinned states have nematic ordering
and the driven states form moving smectic A phases.
A more ordered smectic state containing fewer dislocations appears
for driving along the soft anisotropy direction compared with driving
along the hard anisotropy direction.
When we cycle the drive,
we observe
hysteresis in the dynamics. Specifically,
once the smectic state has formed, it can persist down to 
lower drives than those at which it appeared
on the initial application.
We also find that as the anisotropy increases,
the system is generally less ordered in the high drive states since
dislocations have a lower formation energy.
We map out the dynamic phase diagram for this system
as a function of
varied anisotropy and disorder strength using as our characterization
tools
the orientation of dislocations
and the features in the structure factor. 
Our results should be general to the broader class of driven systems
with two-fold anisotropy driven over random disorder,
which includes electronic liquid crystals, colloidal particles, and
magnetic skyrmions.

%%% ACKNOWLEDGEMENTS
\section*{Acknowledgements}
We are grateful to D.~Minogue, M.~W.~Olszewski and D.~Spulber for assistance with the molecular dynamics simulations and analysis.
This research was supported in part by the Notre Dame Center for Research Computing.
Work at the University of Notre Dame (ER, MRE: MD simulations, data analysis) was supported by the US Department of Energy, Office of Basic Energy Sciences, under Award No. DE-SC0005051.
Part of this work (CR, CJOR: code development) was carried out under support by the US Department of Energy through
the Los Alamos National Laboratory.  Los Alamos National Laboratory is operated by Triad National Security, LLC, for the National Nuclear Security Administration of the U. S. Department of Energy (Contract No. 892333218NCA000001).

%%% BIBLIOGRAPHY
\section*{References}
\bibliographystyle{iopart-num}
\bibliography{mybib}

\end{document}